\documentclass{ieeeaccess}
\usepackage{cite}
\usepackage{amsmath,amssymb,amsfonts}
\usepackage{algorithmic}
\usepackage{graphicx}
\usepackage{textcomp}
\usepackage[acronyms,nonumberlist,nopostdot,nomain,nogroupskip]{glossaries}
\usepackage[font={sf,scriptsize},           % new
            labelfont={bf,color=accessblue},% new
            caption=false]                  % new
           {subfig} 
\usepackage{multirow}

\def\BibTeX{{\rm B\kern-.05em{\sc i\kern-.025em b}\kern-.08em
    T\kern-.1667em\lower.7ex\hbox{E}\kern-.125emX}}
\newcommand{\virg}[1] {``{{#1}}''}

\newacronym{3gpp}{3GPP}{3rd Generation Partnership Project}
\newacronym{adc}{ADC}{Analog to Digital Converter}
\newacronym{5g}{5G}{5th generation}
\newacronym{6g}{6G}{6th generation}
\newacronym{aimd}{AIMD}{Additive Increase Multiplicative Decrease}
\newacronym{amc}{AMC}{Adaptive Modulation and Coding}
\newacronym{aqm}{AQM}{Active Queue Management}
\newacronym{awgn}{AWGN}{Additive White Gaussian Noise}
\newacronym{balia}{BALIA}{Balanced Link Adaptation}
\newacronym{bdp}{BDP}{Bandwidth-Delay Product}
\newacronym{qos}{QoS}{Quality of Service}
\newacronym{pqos}{PQoS}{predictive quality of service}
\newacronym{bf}{BF}{Beamforming}
\newacronym{cc}{CC}{Congestion Control}
\newacronym{cdf}{CDF}{Cumulative Distribution Function}
\newacronym{cn}{CN}{Core Network}
\newacronym{cqi}{CQI}{Channel Quality Information}
\newacronym{cp}{CP}{Control Plane}
\newacronym{csirs}{CSI-RS}{Channel State Information - Reference Signal}
\newacronym{dc}{DC}{Dual Connectivity}
\newacronym{dce}{DCE}{Direct Code Execution}
\newacronym{dci}{DCI}{Downlink Control Information}
\newacronym{dl}{DL}{downlink}
\newacronym{dmr}{DMR}{Deadline Miss Ratio}
\newacronym{dmrs}{DMRS}{DeModulation Reference Signal}
\newacronym{e2e}{E2E}{end-to-end}
\newacronym{ecn}{ECN}{Explicit Congestion Notification}
\newacronym{edf}{EDF}{Earliest Deadline First}
\newacronym{enb}{eNB}{evolved Node Base}
\newacronym{epc}{EPC}{Evolved Packet Core}
\newacronym{es}{ES}{Edge Server}
\newacronym{fdma}{FDMA}{Frequency Division Multiple Access}
\newacronym{fdd}{FDD}{Frequency Division Duplexing}
\newacronym[firstplural=Radio Access Technologies (RATs)]{rat}{RAT}{Radio Access Technology}
\newacronym{fs}{FS}{Fast Switching}
\newacronym{ftp}{FTP}{File Transfer Protocol}
\newacronym{gnb}{gNB}{Next Generation Node Base}
\newacronym{harq}{HARQ}{Hybrid Automatic Repeat reQuest}
\newacronym{hetnet}{HetNet}{Heterogeneous Network}
\newacronym{hh}{HH}{Hard Handover}
\newacronym{hol}{HOL}{Head-of-Line}
\newacronym{ia}{IA}{Initial Access}
\newacronym{ieee}{IEEE}{Institute of Electrical and Electronics Engineers}
\newacronym{imt}{IMT}{International Mobile Telecommunication}
\newacronym{iot}{IoT}{Internet of Things}
\newacronym{ldpc}{LDPC}{Low-Density Parity Check}
\newacronym{los}{LOS}{line-of-sight}
\newacronym{lte}{LTE}{Long Term Evolution}
\newacronym{m2m}{M2M}{Machine to Machine}
\newacronym{ml}{ML}{Machine Learning}
\newacronym{mac}{MAC}{Medium Access Control}
\newacronym{mc}{MC}{Multi-Connectivity}
\newacronym{mcs}{MCS}{Modulation and Coding Scheme}
\newacronym{mec}{MEC}{Mobile Edge Cloud}
\newacronym{mi}{MI}{Mutual Information}
\newacronym{mimo}{MIMO}{Multiple Input Multiple Output}
\newacronym{mmwave}{mmWave}{millimeter wave}
\newacronym{mptcp}{MPTCP}{Multipath TCP}
\newacronym{mr}{MR}{Maximum Rate}
\newacronym{mss}{MSS}{Maximum Segment Size}
\newacronym{mtd}{MTD}{Machine-Type Device}
\newacronym{mtu}{MTU}{Maximum Transmission Unit}
\newacronym{nfv}{NFV}{Network Function Virtualization}
\newacronym{nlos}{NLOS}{non-line-of-sight}
\newacronym{nlosv}{NLOSv}{Vehicle Non-Line-of-Sight}
\newacronym{ofdm}{OFDM}{Orthogonal Frequency Division Multiplexing}
\newacronym{pdcch}{PDCCH}{Physical Downlink Control Channel}
\newacronym{pdcp}{PDCP}{Packet Data Convergence Protocol}
\newacronym{pdsch}{PDSCH}{Physical Downlink Shared Channel}
\newacronym{pdu}{PDU}{Protocol Data Unit}
\newacronym{pf}{PF}{Proportional Fair}
\newacronym{pgw}{PGW}{Packet Gateway}
\newacronym{phy}{PHY}{Physical}
\newacronym{pbch}{PBCH}{Physical Broadcast Channel}
\newacronym[plural=\gls{mme}s,firstplural=Mobility Management Entities (MMEs)]{mme}{MME}{Mobility Management Entity}
\newacronym{prb}{PRB}{Physical Resource Block}
\newacronym{pss}{PSS}{Primary Synchronization Signal}
\newacronym{pscch}{PSCCH}{Physical Sidelink Control Channel}
\newacronym{pucch}{PUCCH}{Physical Uplink Control Channel}
\newacronym{pusch}{PUSCH}{Physical Uplink Shared Channel}
\newacronym{rach}{RACH}{Random Access Channel}
\newacronym{ran}{RAN}{Radio Access Network}
\newacronym{red}{RED}{Random Early Detection}
\newacronym{rf}{RF}{radio frequency}
\newacronym{rlc}{RLC}{Radio Link Control}
\newacronym{rlf}{RLF}{Radio Link Failure}
\newacronym{rrc}{RRC}{Radio Resource Control}
\newacronym{rrm}{RRM}{Radio Resource Management}
\newacronym{rr}{RR}{Round Robin}
\newacronym{rs}{RS}{Remote Server}
\newacronym{rsrp}{RSRP}{Reference Signal Received Power}
\newacronym{rss}{RSS}{Received Signal Strength}
\newacronym{rtt}{RTT}{Round Trip Time}
\newacronym{rw}{RW}{Receive Window}
\newacronym{rx}{RX}{Receiver}
\newacronym{sa}{SA}{standalone}
\newacronym{sack}{SACK}{Selective Acknowledgment}
\newacronym{sap}{SAP}{Service Access Point}
\newacronym{sc}{SC}{Single Carrier}
\newacronym{sch}{SCH}{Secondary Cell Handover}
\newacronym{scoot}{SCOOT}{Split Cycle Offset Optimization Technique}
\newacronym{sdma}{SDMA}{Spatial Division Multiple Access}
\newacronym{sinr}{SINR}{Signal to Interference plus Noise Ratio}
\newacronym{sl}{SL}{Sidelink}
\newacronym{sm}{SM}{Saturation Mode}
\newacronym{snr}{SNR}{Signal-to-Noise-Ratio}
\newacronym{son}{SON}{Self-Organizing Network}
\newacronym{ss}{SS}{Synchronization Signal}
\newacronym{srs}{SRS}{Sounding Reference Signal}
\newacronym{sss}{SSS}{Secondary Synchronization Signal}
\newacronym{tb}{TB}{Transport Block}
\newacronym{tcp}{TCP}{Transmission Control Protocol}
\newacronym{tdd}{TDD}{Time Division Duplexing}
\newacronym{tdma}{TDMA}{Time Division Multiple Access}
\newacronym{tfl}{TfL}{Transport for London}
\newacronym{tm}{TM}{Transparent Mode}
\newacronym{trp}{TRP}{Transmitter Receiver Pair}
\newacronym{tti}{TTI}{Transmission Time Interval}
\newacronym{ttt}{TTT}{Time-to-Trigger}
\newacronym{tx}{TX}{Transmitter}
\newacronym{ue}{UE}{User Equipment}
\newacronym{ul}{UL}{uplink}
\newacronym{uml}{UML}{Unified Modeling Language}
\newacronym{um}{UM}{Unacknowledged Mode}
\newacronym{utc}{UTC}{Urban Traffic Control}
\newacronym{vm}{VM}{Virtual Machine}
\newacronym{rsrq}{RSRQ}{Reference Signal Received Quality}
\newacronym{rssi}{RSSI}{Received Signal Strength Indicator}
\newacronym{crs}{CRS}{Cell Reference Signal}
\newacronym{nsa}{NSA}{Non Stand Alone}
\newacronym{mrdc}{MR-DC}{Multi \gls{rat} \gls{dc}}
\newacronym{endc}{EN-DC}{E-UTRAN-\gls{nr} \gls{dc}}
\newacronym{5gc}{5GC}{5G Core}
\newacronym{si}{SI}{Study Item}
\newacronym{iab}{IAB}{Integrated Access and Backhaul}
\newacronym{wf}{WF}{Wired-first}
\newacronym{hqf}{HQF}{Highest-quality-first}
\newacronym{pa}{PA}{Position-aware}
\newacronym{mlr}{MLR}{Maximum-local-rate}
\newacronym{wbf}{WBF}{Wired Bias Function}
\newacronym{mib}{MIB}{Master Information Block}
\newacronym{sib}{SIB}{Secondary Information Block}
\newacronym{rnti}{RNTI}{Radio Network Temporary Identifier}
\newacronym{dft}{DFT}{Discrete Fourier Transform}
\newacronym{kpi}{KPI}{Key Performance Indicator}
\newacronym{ppp}{PPP}{Poisson Point Process}
\newacronym{v2v}{V2V}{Vehicle-to-Vehicle}
\newacronym{wave}{WAVE}{Wireless Access in Vehicular Environments}
\newacronym{udp}{UDP}{User Datagram Protocol}
\newacronym{upa}{UPA}{Uniform Planar Array}
\newacronym{fec}{FEC}{Forward Error Correction}
\newacronym{v2x}{V2X}{Vehicle-To-Everything}
\newacronym{psfch}{PSFCH}{Physical Sidelink Feedback Channel}
\newacronym{pssch}{PSSCH}{Physical Sidelink Shared Channel}
\newacronym{csma}{CSMA}{Carrier Sense Multiple Access}
\newacronym{v2n}{V2N}{Vehicle-to-Network}
\newacronym{wlan}{WLAN}{Wireless Local Area Network}
\newacronym{cav}{CAV}{Connected and Autonomous Vehicle}
\newacronym{v2i}{V2I}{Vehicle-to-Infrastructure}
\newacronym{d2d}{D2D}{Device-to-Device}
\newacronym{c-its}{C-ITS}{Connected Intelligent Transportation System}
\newacronym{fr2}{FR2}{Frequency Range 2}
\newacronym{fr1}{FR1}{Frequency Range 1}
\newacronym{bs}{BS}{Base Station}
\newacronym{sdu}{SDU}{Service Data Unit}
\newacronym{csi}{CSI}{Channel State Information}
\newacronym{scs}{SCS}{Subcarrier Spacing}
\newacronym{sumo}{SUMO}{Simulation of Urban MObility}
\newacronym{prr}{PRR}{Packet Reception Ratio}
\newacronym{edca}{EDCA}{Enhanced Distribution Channel Access}
\newacronym{sdap}{SDAP}{Service Data Adaptation Protocol}
\newacronym{iiot}{IIoT}{Industrial Internet of Things}
\newacronym{agv}{AGV}{Automated Guided Vehicle}
\newacronym{cm}{C/M}{Controller/Master}
\newacronym{soa}{SoA}{State-of-the-Art}
\newacronym{snpn}{SNPN}{Standalone Non-Public Network}
\newacronym{pninpn}{PNI-NPN}{Public Network Interface Non-Public Network}
\newacronym{urllc}{URLLC}{Ultra-Reliable Low-Latency Communication}
\newacronym{embb}{eMBB}{enhanced Mobile BroadBand}
\newacronym{ai}{AI}{Artificial Intelligence}
\newacronym{mab}{MAB}{Multi-Armed Bandit}
\newacronym{su}{SU}{Scheduling Unit}
\newacronym{ra}{RA}{Random Agent}
\newacronym{na}{NA}{Neural Agent}
\newacronym{ucba}{UCB-A}{UCB Agent}
\newacronym{tsa}{TS-A}{Thompson Sampling Agent}
\newacronym{ucb}{UCB}{Upper Confidence Bound}
\newacronym{ts}{TS}{Thompson Sampling}
\newacronym{inf}{InF}{Indoor Factory}
\newacronym{infsl}{InF-SL}{Indoor Factory - Sparse Clutter, Low BS}
\newacronym{infdl}{InF-DL}{Indoor Factory - Dense Clutter, Low BS}
\newacronym{infsh}{InF-SH}{Indoor Factory - Sparse Clutter, High BS}
\newacronym{infdh}{InF-DH}{Indoor Factory - Dense Clutter, High BS}
\newacronym{us}{US}{Uplink Scheduler}
\newacronym{nn}{NN}{Neural Network}
\newacronym{das}{DAS}{Distributed Antenna System}
\newacronym{rb}{RB}{Resource Block}
\newacronym{rl}{RL}{Reinforcement Learning}
\newacronym{uav}{UAV}{Unmanned Aerial Vehicle}
\newacronym{5gacia}{5G-ACIA}{5G Alliance for Connected Industries and Automation}
\newacronym{fl}{FL}{Federated Learning}
\newacronym{ack}{ACK}{Acknowledgment} 
\newacronym{dnn}{DNN}{Deep Neural Network}
\newacronym{sr}{SR}{Scheduling Request}
\newacronym{bsr}{BSR}{Buffer Status Report}
\newacronym{nr}{NR}{New Radio}
\newacronym{cb}{CB}{Contention-Based}
\newacronym{cf}{CF}{Contention-Free}
\newacronym{ce}{CE}{Control Element}
\newacronym{sps}{SPS}{Semi-Persistent Scheduling}
\newacronym{ds}{DS}{Dynamic Scheduling}
\newacronym{xr}{XR}{Extended Reality}
\newacronym{ofdma}{OFDMA}{Orthogonal Frequency Division Multiple Access}
\newacronym{cdma}{CDMA}{Code Division Multiple Access}
\newacronym{noma}{NOMA}{Non-Orthogonal Multiple Access}
\newacronym{ibi}{IBI}{Instantaneous Buffer Information}
\newacronym{am}{AM}{Acknowledge Mode}
\begin{document}
\history{Date of publication xxxx 00, 0000, date of current version xxxx 00, 0000.}
\doi{xxxx}

\title{Analysis of a contention-based approach over 5G NR for Federated Learning in an Industrial Internet of Things scenario}
\author{\uppercase{Giampaolo Cuozzo}\authorrefmark{1}, \uppercase{Jonas Pettersson}\authorrefmark{2},
\uppercase{and Massimo Condoluci}\authorrefmark{3}}
\address[1]{National Laboratory of Wireless Communications (WiLab), CNIT, Bologna, Italy (email: giampaolo.cuozzo@cnit.it)}
\address[2]{Ericsson Research, Luleå, Sweden (email: jonas.pettersson@ericsson.com)}
\address[3]{Ericsson Research, Stockholm, Sweden (email: massimo.condoluci@ericsson.com)}
% \tfootnote{This paragraph of the first footnote will contain support 
% information, including sponsor and financial support acknowledgment. For 
% example, ``This work was supported in part by the U.S. Department of 
% Commerce under Grant BS123456.''}

\markboth
{G. Cuozzo \headeretal: Analysis of a contention-based approach over 5G NR for Federated Learning in an Industrial Internet of Things scenario}
{G. Cuozzo \headeretal: Analysis of a contention-based approach over 5G NR for Federated Learning in an Industrial Internet of Things scenario}

\corresp{Corresponding author: G. Cuozzo (e-mail: giampaolo.cuozzo@cnit.it).}

\begin{abstract}
The growing interest in new applications involving co-located heterogeneous requirements, such as the Industrial Internet of Things (IIoT) paradigm, poses unprecedented challenges to the uplink wireless transmissions. Dedicated scheduling has been the fundamental approach used by mobile radio systems for uplink transmissions, where the network assigns contention-free resources to users based on buffer-related information. The usage of contention-based transmissions was discussed by the 3rd Generation Partnership Project (3GPP) as an alternative approach for reducing the uplink latency characterizing dedicated scheduling. Nevertheless, the contention-based approach was not considered for standardization in LTE due to limited performance gains. However, 5G NR introduced a different radio frame which could change the performance achievable with a contention-based framework, although this has not yet been evaluated. This paper aims to fill this gap. We present a contention-based design introduced for uplink transmissions in a 5G NR IIoT scenario. We provide an up-to-date analysis via near-product 3GPP-compliant network simulations of the achievable application-level performance with simultaneous Ultra-Reliable Low Latency Communications (URLLC) and Federated Learning (FL) traffic, where the contention-based scheme is applied to the FL traffic. The investigation also involves two separate mechanisms for handling retransmissions of lost or collided transmissions. Numerical results show that, under some conditions, the proposed contention-based design provides benefits over dedicated scheduling when considering FL upload/download times, and does not significantly degrade the performance of URLLC.
\end{abstract}


\titlepgskip=-15pt

\maketitle

%%%%%%%%%%%%%%%%%%%%%%%%%%%%%%%%%%%%%%%%%%%%%%
%%%%%%%%%%%%%%%%%%%%%%%%%%%%%%%%%%%%%%%%%%%%%%
\section{Introduction}
\label{sec:intro}

Next-generation mobile radio networks will support new use cases and, consequently, new traffic types \cite{metis5G, navarro2020survey, chettri2019comprehensive, erunkulu20215g, elsayed2019ai}. One exemplary emerging application is the \gls{iiot}, where wireless technologies ensure the interconnection between industrial assets (e.g., valves, pumps, robotic arms, etc.) and the control rooms of industry plants \cite{5GIndustrie4zero, 5gaciaarchitecture} to realize digital twins of physical industrial entities, promote \gls{xr}-based maintenance operations, or 
support distributed \gls{ml} frameworks such as \gls{fl} \cite{hexax2021, giordani2020toward, jiang2021road, tataria20216g, tao2018digital, ganjalizadeh2022interplay}.
% The support of new traffic types over mobile radio networks allows the realization of new use cases \cite{metis5G, navarro2020survey, chettri2019comprehensive, erunkulu20215g, elsayed2019ai}. 
% One example is industrial automation, a target use case for \gls{5g} which leverages the support of \gls{urllc} for interconnecting industrial assets (e.g., valves, pumps, robotic arms, etc..) with the control room \cite{5GIndustrie4zero, 5gaciaarchitecture}. Nevertheless, some applications may demand technological changes towards future \gls{6g} networks \cite{hexax2021, giordani2020toward, jiang2021road, tataria20216g}, such as the realization of digital twins and/or collaborative robots in industrial environments, where humans could interact with digital representations of industrial assets using \gls{xr} \cite{tao2018digital} and devices could autonomously cooperate using distributed \gls{ml} frameworks (e.g., \gls{fl} \cite{ganjalizadeh2022interplay}).  

Notably, the main characteristic of these new data transfers is that they put more effort into the uplink direction, whereas legacy traffics, such as web browsing, are rather downlink-heavy. For instance, uplink performance is as important as downlink for fast convergence of \gls{fl} algorithm, where devices upload the results of their local training to a central entity (upstream) which performs aggregation and re-distributes the updated model (downstream) until all nodes utilize the same version \cite{3gpp22874}. In this regard, the literature has been investigating several approaches to optimize the uplink data transmissions that mainly belong to two categories: \gls{cf} and \gls{cb}. According to the former, \glspl{ue} transmit via dedicated radio resources that can be either time slots (\gls{tdma}) \cite{huang2012evolution, hadded2015tdma}, frequency channels (\gls{fdma}) \cite{zhang2015advances}, or their combination \cite{abu2013uplink, salem2010overview, yaacoub2011survey}, as well as orthogonal spreading codes in a \gls{cdma} approach \cite{gilhousen1991capacity, kusume2011idma, buratti2022ocdma}, and spatial beams in a \gls{mimo} network \cite{ahmed2018survey, choi2007opportunistic}. As for the \gls{cb} uplink transmissions, besides the proliferation of well-known studies on ALOHA-based solutions and \gls{csma} protocols \cite{akkarajitsakul2011game, zheng2014investigation, lam2006polling, bianchi2000performance,ghazvini2012game}, a recent hot topic is called \gls{noma}, where smart receivers are designed to mitigate the interference produced by uplink transmissions that exploit the same radio resource \cite{dai2018survey, ding2017survey}. \newline 
From a standardization viewpoint, the \gls{3gpp} has been considering dedicated scheduling as the main approach for  uplink data transmission, with the network assigning dedicated radio resources (grants) upon receiving explicit requests from each \gls{ue}. Radio resources could be either granted in a dynamic way based on the amount of data a \gls{ue} has in its buffer or could be allocated in a semi-persistent way with an allocation repeating over a certain amount of time.
The usage of \gls{cb} approach has been studied for \gls{lte} to allow \gls{ue}s to directly transmit data in uplink without having to wait for a dedicated grant \cite{3gppR2_093812_CB}. Nevertheless, performance gains of \gls{cb} over \gls{lte} were limited and achievable only in scenarios with low load and small-size uplink data, hence standardization continued to focus on dedicated scheduling as the main approach for uplink data transmission. 

However, with the proliferation of new uplink-oriented applications with heterogenous requirements, there is a renewed interest in exploring the potential benefits of \gls{cb} designs for \gls{3gpp}-compliant networks. Additionally, \gls{5g} \gls{nr} foresees substantial differences w.r.t \gls{lte} that might really unleash the potential benefits of \gls{cb} schemes. Therefore, the aim of this paper is to re-visit the work done by \gls{3gpp} and to give a first assessment of the achievable performance of \gls{cb} uplink transmissions applied to \gls{5g} \gls{nr}. We present a \gls{cb} design for 5G NR \gls{pusch}, and we consider different mechanisms for handling retransmissions of lost or collided transmissions. Unlike previous assessments done by \gls{3gpp}, we consider extensive network simulations to assess the application-level performance achieved by \gls{fl} traffic in an \gls{iiot} scenario when using the proposed \gls{cb} design for \gls{5g} \gls{nr} \gls{pusch}, focusing on both upstream and downstream performance. Numerical results show that the considered \gls{cb} design for \gls{5g} \gls{nr} \gls{pusch} provides benefits over dedicated scheduling under some conditions, and scales well with the number of \glspl{ue}, by also poorly deteriorating the application-level performance of other higher-priority traffic flows.
    
The paper is structured as follows. In Sec.~\ref{sec:related_works} we clarify the original contributions of this paper by reviewing both the academic literature and \gls{3gpp} standards. Sec.~\ref{sec:contention_based_scheduling} describes the considered \gls{cb} design for \gls{nr} \gls{pusch}, whereas Secs.~\ref{sec:system_model} and \ref{sec:performance_metrics} present the system model and the metrics used for the performance evaluation. Finally, in Sec.~\ref{sec:performance_evaluation} we present the corresponding numerical results, while in Sec.~\ref{sec:conclusions} we summarize the main achievements and possible future works.

%%%%%%%%%%%%%%%%%%%%%%%%%%%%%%%%%%%%%%%%%%%%%%
%%%%%%%%%%%%%%%%%%%%%%%%%%%%%%%%%%%%%%%%%%%%%%
\section{Related Works}
\label{sec:related_works}

%%%%%%%%%%%%%%%%%%%%%%%%%%%%%%%%%%%%%%%%%%%%%%
\subsection{Literature review on uplink data transmissions}
The academic literature analyzes several approaches to shrink the uplink latency provided by dedicated scheduling, where \glspl{ue} willing to transmit data have to first request radio resources from the network. Some works propose improvements of the semi-persistent allocation mechanisms \cite{arnjad2018latency}, where the network reserves a given number of dedicated radio resources for a limited amount of time. In this regard, the authors in \cite{feng2019predictive} study predictive algorithms for the radio resource assignments by considering an \gls{lte} network, whereas the potential benefits of a traffic-aware semi-persistent scheduler are investigated in \cite{cuozzo2022enabling} for a private 5G NR network tailored to an \gls{iiot} environment. Semi-persistent resource allocations reduce the control plane overhead, but fail in managing unpredictable/highly-variable traffic and do not scale well with the offered traffic due to an intrinsic spectral inefficiency. 

To overcome the above limitations, the literature is proposing grant-free transmissions \cite{ding2021enabling, shahab2020grant}, that is, a distributed scheme where \glspl{ue} can autonomously select the radio resources to be used for their uplink transmission without relying on any grant reception, thereby introducing possible collisions. This approach is tailored to aperiodic (or uncertain) traffic but its \gls{cb} nature undermines communication reliability. Some works \cite{srinath2022grant, berardinelli2018reliability, liu2020analyzing, lucas2019capacity, jacobsen2017system} try to mitigate the collision impact by studying both, the optimal number of a-priori packet duplications and how to manage the acknowledgments of the duplicates, leading to the consequent trade-off between resource efficiency and reliability. Conversely, others investigate sensing mechanisms and/or interference cancellation techniques \cite{lucas2022sensing, mahmood2019uplink, au2014uplink}, as well as considering \glspl{ue} that leverage \gls{ml} to learn how to optimally select the radio resources based on their past experience \cite{pase2022distributed}. However, distributed solutions imply a higher complexity at the \gls{ue}-side which may be unfeasible in some scenarios (e.g., for \gls{iiot} applications), and their optimality applies only to particular cases.
%%%%%%%%%%%%%%%%%%%%%%%%%%%%%%%%%%%%%%%%%%%%%%
\subsection{Standardization review on uplink data transmissions}
\label{sec:uplink_data_transmissions_3gpp}

\gls{ds} is the main approach used in \gls{3gpp}-compliant networks to support the transmission of uplink data with variable size and no periodic patterns \cite{parkvall20185g}. 
Fig.~\ref{fig:dynamic_scheduling} shows the timing diagram of \gls{ds}. First, a \gls{ue} with no allocated grants (i.e., dedicated radio resources) waits for an occasion to send a \gls{sr} to indicate to the \gls{gnb} that it has new data  to be sent, then the \gls{gnb} replies with a grant (\virg{Grant\#1} in Fig.~\ref{fig:dynamic_scheduling}) containing the set of radio resources that are allocated for the first uplink transmission. Consequently, the \gls{ue} will create a \gls{tb} (i.e., \gls{mac} \gls{pdu}) based on the received grant. The \gls{tb} will be used to carry (i) the \gls{bsr}, i.e., a \gls{mac} \gls{ce} indicating the number of bytes left in its transmission buffer, and (ii) any data that may fit into it\footnote{\glspl{sr}, grants, and the message containing \glspl{bsr} plus data are mapped to \glspl{pucch}, \glspl{pdcch} and \glspl{pusch}, respectively.}. 
%Hence, depending on the amount of available radio resources,
The number of resources allocated by the first grant could be enough to allow the \gls{ue} to transmit all data in its queue, but this cannot be guaranteed as the \gls{gnb} has not yet information on how much data the \gls{ue} has to send. Hence, depending on the received \gls{bsr}, the \gls{gnb} could send one or more new grants to allow the \gls{ue} to free up its buffer (transmissions highlighted with a dashed box in Fig.~\ref{fig:dynamic_scheduling}). \newline
As a matter of fact, \gls{ds} is a very flexible approach because it allows tailoring the radio resources allocated to a \gls{ue} based on its buffer status and cell load, as well as adjusting transmission parameters (e.g., \gls{mcs}) based on its channel quality. Nevertheless, the interval from when new data reaches \gls{ue}'s buffer to when the \gls{gnb} knows how much data the \gls{ue} has actually to send is not negligible and this impacts the overall uplink latency performance.

\Figure[t!](topskip=0pt, botskip=0pt, midskip=0pt)[width=3.25in]{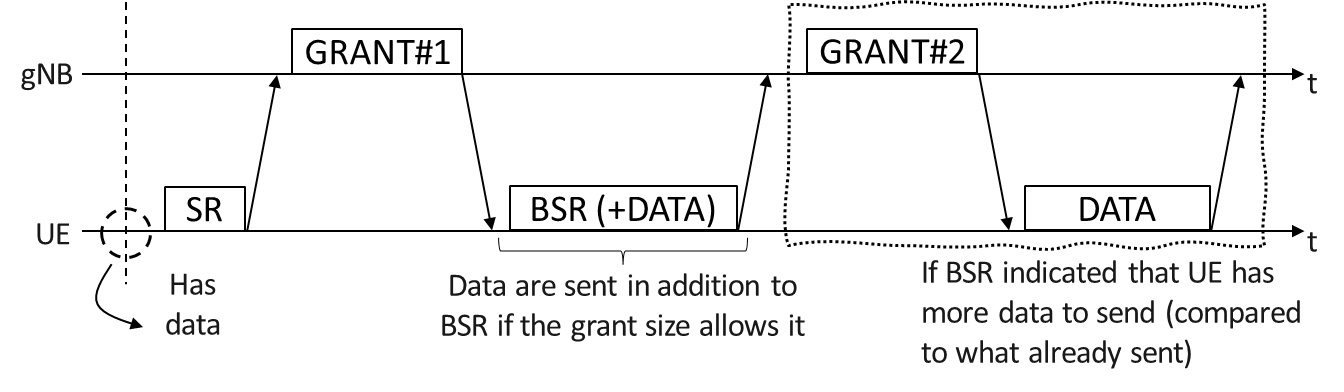}
 {High-level time diagram of the basic dynamic scheduling principle.\label{fig:dynamic_scheduling}}

The \gls{3gpp} studied possible uplink latency reduction techniques for \gls{lte} in Rel. 9 \cite{3gpp36912_TR_enh_LTEA, 3gppR2_093812_CB} and in Rel. 14. \cite{3gpp36881_TR_latency_red, 3gppR2_154122_CB_noDRMS,3gppR2_154191_CB_wDRMS}. For \gls{ds}, it was proposed to increase the frequency of \gls{sr} occasions to reduce the first component of uplink delay.
Other solutions were based on \gls{sps}, with periodic fixed-size allocation dedicated to a \gls{ue} which would allow a \gls{ue} to directly start transmitting its buffered data. However, since the \gls{ue} could have no data to transmit in a given \gls{sps} occasion, it could do padding or skip the transmission opportunity depending on the configuration sent by the \gls{gnb}. 
Solutions \cite{3gppR2_093812_CB, 3gppR2_154122_CB_noDRMS,3gppR2_154191_CB_wDRMS} were instead based on the usage of a \gls{cb} \gls{pusch}, where a \gls{ue} could directly transmit its uplink data using a pre-configured \gls{pusch} allocation which is shared among multiple \gls{ue}s, thereby introducing collisions in the network.
Regarding handling of collisions, the proposal \cite{3gppR2_154122_CB_noDRMS} considered that the \gls{gnb} could not distinguish among colliding \gls{ue}s. A colliding \gls{ue} will not receive any feedback (acknowledgment of successful reception) by the \gls{gnb} and this will trigger a retransmission. In this scheme, the \gls{ue} will perform backoff when selecting the next \gls{cb} \gls{pusch} occasion for transmission.
The proposal \cite{3gppR2_154191_CB_wDRMS}, instead, considered that the \gls{gnb} could distinguish colliding \gls{ue}s through \gls{dmrs}-based \gls{ue} identification \cite{parkvall20185g, huang2015reference}. In this way, the \gls{gnb} can, at least, acquire knowledge of which \gls{ue}s collided and consequently schedule a dedicated \gls{pusch} resource for their retransmission (thus avoiding further collisions).
Moreover, the study in \cite{3gppR2_093812_CB} provided an analysis of achievable performance when using \gls{cb} for traffic upload and download and considering one \gls{ue}, whereas relationships between uplink load, collision probability, and uplink latency characterizing the aforementioned solutions can be found in \cite{3gppR2_154122_CB_noDRMS,3gppR2_154191_CB_wDRMS}. By considering these works, the calculations in \cite{3gpp36881_TR_latency_red, 3gppR2_156402_CB_wDRMS_analys} highlight that the uplink delay for \gls{cb} \gls{pusch} transmissions is difficult to be kept stable if the collision probability (which depends on how many \gls{ue}s share the same \gls{cb} allocation) becomes too high, whereas solutions based on \gls{sr} frequency increase and on \gls{sps} allow a more predictable delay performance at the expense of a reduced uplink capacity. Consequently, 
\cite{3gpp36881_TR_latency_red, 3gpp36912_TR_enh_LTEA} concluded that the gains of the \gls{cb} \gls{pusch} solutions were too limited for \gls{lte} compared to \gls{ds} or \gls{sps} to motivate the required extra standardization work.

\subsection{Original contributions of this paper}
Besides the degradation of performance caused by collision, the study from \gls{3gpp} did not provide an exhaustive analysis of the behavior of \gls{cb} over \gls{lte}. Furthermore, \gls{5g} \gls{nr} brought different changes compared to \gls{lte} which could influence the achievable performance of a \gls{cb} approach. Overall, the contributions of this paper are: 
\begin{itemize}
    \item Introduce a \gls{cb} design for \gls{5g} \gls{nr} \gls{pusch}.
    \item Analyse performance when legacy \gls{ds} and \gls{cb} for \gls{5g}  \gls{nr} \gls{pusch} are simultaneously used, by also assessing the impact of two different retransmission mechanisms which are inspired from previous \gls{3gpp} studies \cite{3gppR2_154122_CB_noDRMS, 3gppR2_154191_CB_wDRMS}.
    \item Analyse performance of \gls{cb} for \gls{5g} \gls{nr} \gls{pusch} when applied to a \gls{fl}-based \gls{iiot} scenario, where there is a correlation among uplink transmissions of \gls{fl} \gls{ue}s, thus creating a more challenging scenario for \gls{cb}.
    \item Analyse both downstream and upstream flows.
    \item Analyse the trade-off and the relationships among different metrics related to \gls{cb} for \gls{5g} \gls{nr} \gls{pusch}.
\end{itemize}

%%%%%%%%%%%%%%%%%%%%%%%%%%%%%%%%%%%%%%%%%%%%%%
%%%%%%%%%%%%%%%%%%%%%%%%%%%%%%%%%%%%%%%%%%%%%%
\section{CB for NR PUSCH}
\label{sec:contention_based_scheduling}

\begin{figure}[!t]
\centering
 \subfloat[Pessimistic case, where the \gls{ue} has new data to transmit but has not yet received a \gls{cb} grant.]
	{\includegraphics[width=0.9\linewidth]{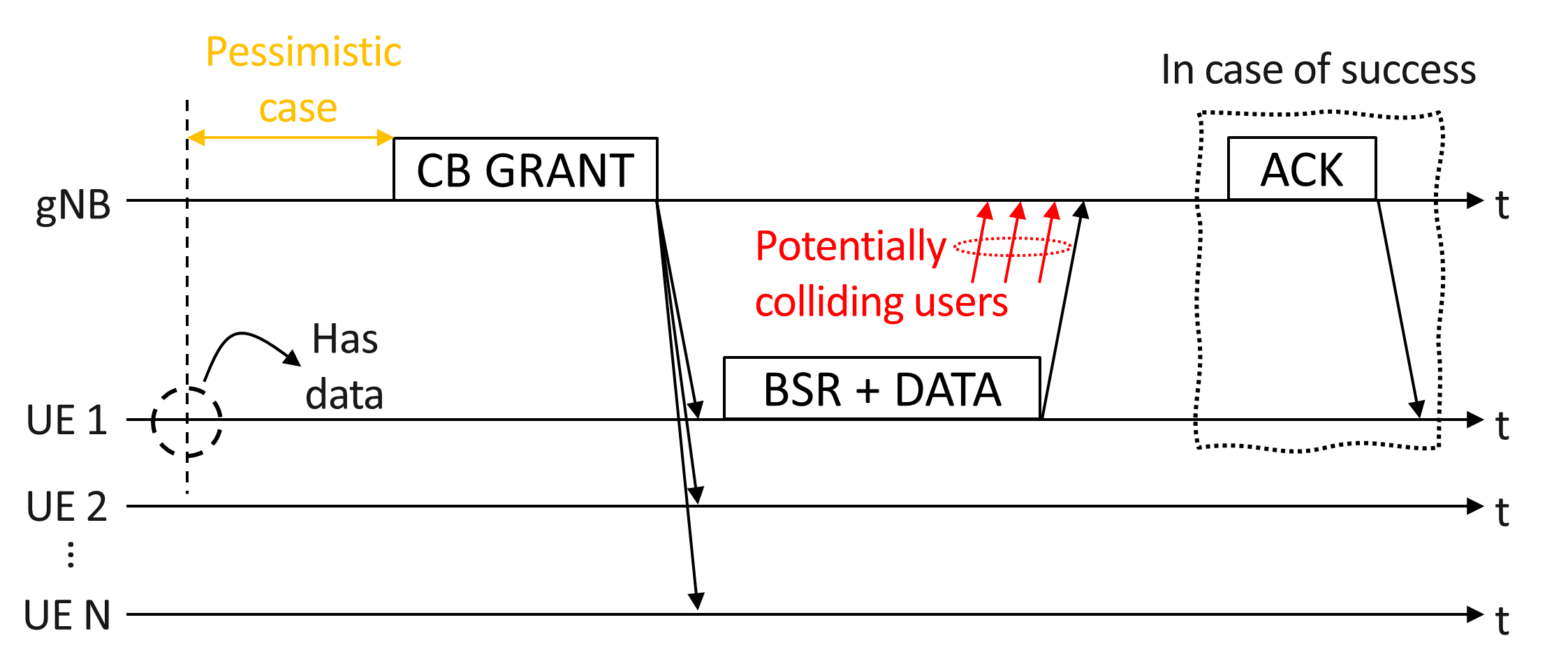}
	\label{fig:timing_diagram_pessimistic}} \\
\subfloat[Optimistic case, where the \gls{ue} has already received a \gls{cb} grant before new data to be transmitted.]
	{\includegraphics[width=0.9\linewidth]{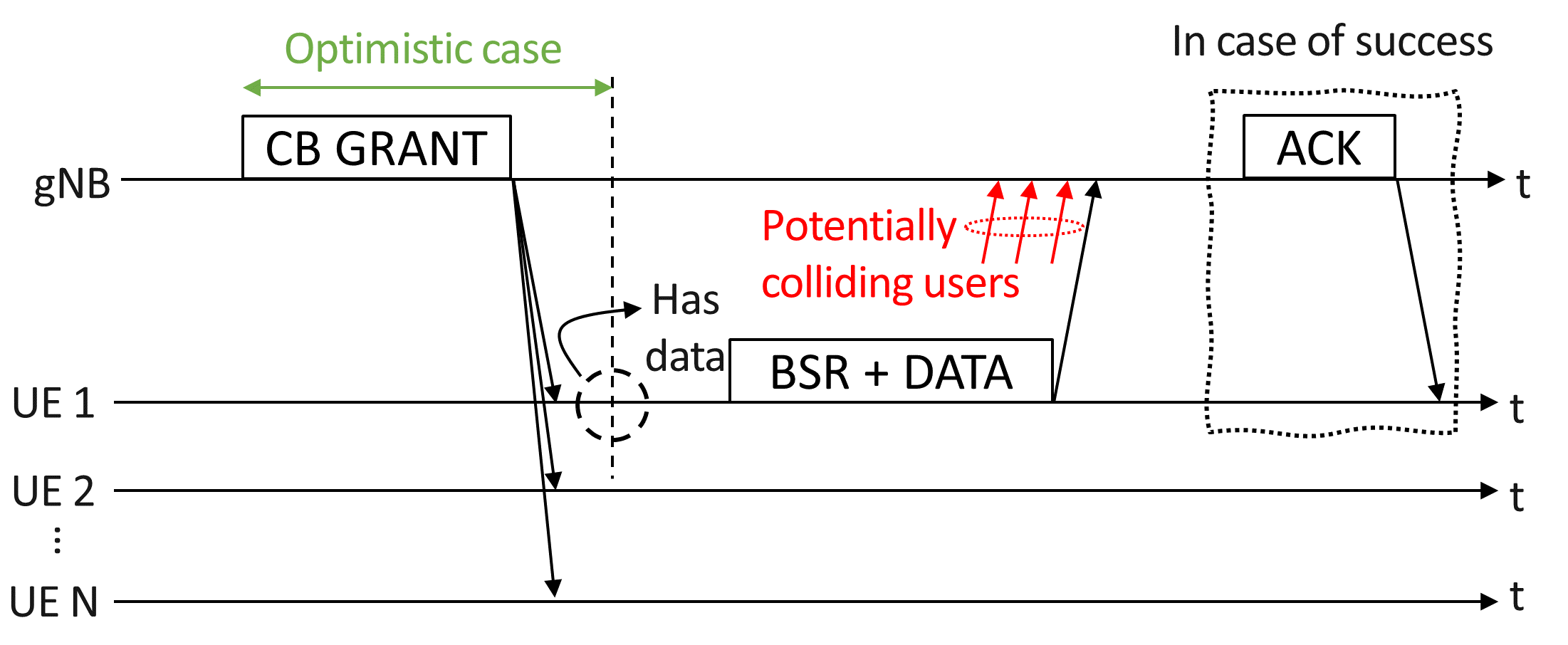}
	\label{fig:timing_diagram_optimistic}}
\caption{Timing diagram of the considered \gls{cb} for \gls{nr} \gls{pusch} design, showing the relationship between reception of \gls{cb} grant and availability of uplink data.}
\label{fig:timing_diagram}
\end{figure}

%%%%%%%%%%%%%%%%%%%%%%%%%%%%%%%%%%%%%%%%%%%%%%%%%%%%%%%%%%%%%%%%%%%%%%%%%%%%%%
\subsection{General 5G radio framework}
The time axis is divided into \emph{slots}, composed of 14 \gls{ofdm} symbols each, whereas the frequency axis is partitioned into \glspl{rb}, that is, sets of 12 subcarriers \cite{parkvall20185g}. We consider an \gls{fdd} scheme where the \gls{gnb} manages the uplink/downlink radio resources, that is, a set of \gls{ofdm} symbols and \glspl{rb} (space/power domains are not considered in this paper).

%%%%%%%%%%%%%%%%%%%%%%%%%%%%%%%%%%%%%%%%%%%%%%%%%%%%%%%%%%%%%%%%%%%%%%%%%%%%%%
\subsection{Contention-based design}
\label{sec:contention_based_design}
The \gls{gnb} allocates a portion of uplink radio resources to a given set of \gls{ue}s. We refer to this allocation as \textit{\gls{cb} grant, \gls{cb} resource, or \gls{cb} allocation}, equivalently. In our current implementation, we consider that a \gls{cb} grant is created at each slot\footnote{Of course, other approaches could be considered, e.g., creating the \gls{cb} grant as a semi-persistent allocation thus avoiding the transmission of \gls{cb} grants at each slot. Nevertheless, we considered this approach for simplicity of implementation, and because it allowed us to analyze scenarios with a dynamic variation of the resources allocated to the \gls{cb} grant.}. The \gls{cb} grant is broadcasted to the \glspl{ue} associated with that \gls{cb} resource and contains the following two main pieces of information:
\begin{enumerate}
\item The time/frequency location and dimension (in terms of number of \gls{ofdm} symbols and \glspl{rb}) of the \gls{cb} resource\footnote{Notice that, in our design, the colliding transmissions are completely overlapped in time and frequency, and this means that the \gls{gnb} can avoid performing blind decoding, that is, blindly searching for possible transmissions within a given time-frequency resource.};
\item The \gls{mcs} associated to the \gls{cb} allocation. In our current implementation, we consider that the 
\gls{gnb} has \glspl{csi}\footnote{Specifically, the \gls{gnb} computes the \gls{csi} upon receiving the periodical \gls{cqi} transmissions made by the \glspl{ue}.} of the \gls{ue}s, hence the \gls{mcs} is chosen according to the \gls{ue} in the worst channel condition. If \glspl{csi} are not available, the most conservative \gls{mcs} is selected. 
% \item The set of \gls{harq} processes, one per \gls{ue}, indicating which \gls{mac} \gls{pdu} each \gls{ue} should transmit via the considered \gls{cb} grant\footnote{Specifically, one \gls{harq} process identifies a single \gls{tb} (or \gls{mac} \gls{pdu}) which is sent to the \gls{phy} layer. Notice that a single \gls{mac} \gls{pdu} can contain several \gls{mac} layer \glspl{sdu} (in case of concatenation).}.
\end{enumerate}

Only the \glspl{ue} with non-empty queues will exploit the \gls{cb} grant to transmit \gls{bsr} and data, whereas the others will ignore it. In case of successful \gls{cb} transmission, the \gls{gnb} replies with a positive \gls{ack}.%\footnote{\glspl{ack} are mapped to \glspl{pdsch}.}.

Fig.~\ref{fig:timing_diagram} shows the timing diagram of the aforementioned \gls{cb} design, focusing on the time relationship between the availability of a \gls{cb} allocation and the presence of data in the \gls{ue}'s buffer.  Fig.~\ref{fig:timing_diagram_pessimistic} depicts the pessimistic case, i.e., the \gls{ue} has new data available at its buffer but has no \gls{cb} resources granted for transmission, so it has to wait to receive a \gls{cb} grant. Indeed, this is possible because the creation of \gls{cb} grants and data are independent events. Fig.~\ref{fig:timing_diagram_optimistic} represents the optimistic case, where the \gls{ue} has already received a \gls{cb} grant and thus new data which reached its buffer can directly be sent over the \gls{cb} resources.

However, a \gls{cb} uplink transmission can fail either due to collisions or link failures. In the former case, we consider that collisions are always harmful, i.e., no capture effect is considered. Nonetheless, regardless of the reason for the missed reception, the \gls{gnb} will not reply with any \gls{ack}, thereby triggering a retransmission \cite{parkvall20185g, 3gppR2_154122_CB_noDRMS}. 

%we consider that the \gls{gnb} sends feedback either in the form of an \gls{ack}\footnote{\glspl{ack} are mapped to \glspl{pdsch}.} to inform the \gls{ue}s which used the \gls{cb} resources about the success or a dedicated grant to inform that a retransmission is needed. 

%%%%%%%%%%%%%%%%%%%%%%%%%%%%%%%%%%%%%%%%%%%%%%%%%%%%%%%%%%%%%%%%%%%%%%%%%%%%%%
\subsection{Retransmission mechanisms}
\label{sec:retx_mechanisms}
By recalling that \gls{5g} \gls{nr} relies on \gls{harq} at the \gls{mac} layer, we defined a maximum number of retransmissions for each \gls{tb}, $N_{\rm RX}$. In particular, when a \gls{ue} wants to retransmit a \gls{tb} after $N_{\rm RX} + 1$ times, the \gls{mac} layer declares a \gls{harq} failure, and an \gls{rlc} retransmission is triggered since we consider \gls{am} \gls{rlc}. 

Specifically, two retransmission mechanisms are considered, where all \glspl{ue} implement the same mechanism within one simulation round.

\subsubsection{Retransmissions on dedicated resources} 
\label{sec:re-tx_dedicated}
In this case, we assume that the \gls{gnb} can retrieve the identity of the colliding \glspl{ue}, and thereby it can reserve dedicated radio resources for each colliding \gls{ue} to retransmit the failed \gls{tb}. In particular, the dedicated grant will indicate which \gls{cb} resource was used by the failed attempt so that the \gls{ue} can know what \gls{tb} to retransmit.

%Just as in the case of ordinary \gls{ds} this is done through a dynamic grant. This grant indicates on which contention based grant the failed attempt was detected so that the \gls{ue} can know what \gls{harq} process to retransmit. 

\textbf{Remark 1}: The \gls{mcs} associated to retransmissions on dedicated resources is no longer dependent on the worst channel conditions but is tailored to the \gls{csi} of the specific \gls{ue} (if available). 

\textbf{Remark 2}: In real-world implementations, the use of orthogonal signals, such as those obtained with a proper mapping of \gls{dmrs} symbols, can let the \gls{gnb} know the identity of the colliding \glspl{ue} \cite{parkvall20185g}. However, these types of signals are usually in a finite number, and this may limit the number of \glspl{ue} that can exploit a \gls{cb} allocation. This aspect is left to future studies, i.e., we have assumed that such a mechanism can distinguish all \glspl{ue} associated with the \gls{cb} resource. 

%In this study we have assumed that such a mechanism can distinguish all \glspl{ue} associated with the \gls{cb} resource. The implementation is left to future studies.  

\subsubsection{Retransmissions on contention-based resources}
\label{sec:re-tx_contention}
In this case, when the \gls{ack} is not received, each \gls{ue} will perform backoff before retransmitting again via \gls{cb} resources. During backoff, each \gls{ue} will stay silent, i.e., it will not exploit any \gls{cb} grant, for a number of slots uniformly distributed over the interval $[0, T_{\rm BO}]$. Remarkably, since retransmissions refer to the same \gls{tb} created for the failed transmission, the \gls{mcs} of the \gls{cb} allocation used for the retransmission may not be compatible with that associated with the failed \gls{tb}, since this depends on how the channel has changed for the worst \gls{ue}. Therefore, when using this retransmission policy, \glspl{ue} do not rely on \gls{harq} retransmissions, that is, we set $N_{\rm RX} = 0$. Conversely, an \gls{rlc} retransmission is triggered directly, thereby generating a new set of \gls{tb}.

%%%%%%%%%%%%%%%%%%%%%%%%%%%%%%%%%%%%%%%%%%%%%%
%%%%%%%%%%%%%%%%%%%%%%%%%%%%%%%%%%%%%%%%%%%%%%
\section{System Model}
\label{sec:system_model}

%%%%%%%%%%%%%%%%%%%%%%%%%%%%%%%%%%%%%%%%%%%%%%
\subsection{Scenario}
\label{sec:scenario}
\Figure[t!](topskip=0pt, botskip=0pt, midskip=0pt)[width=3.25in]{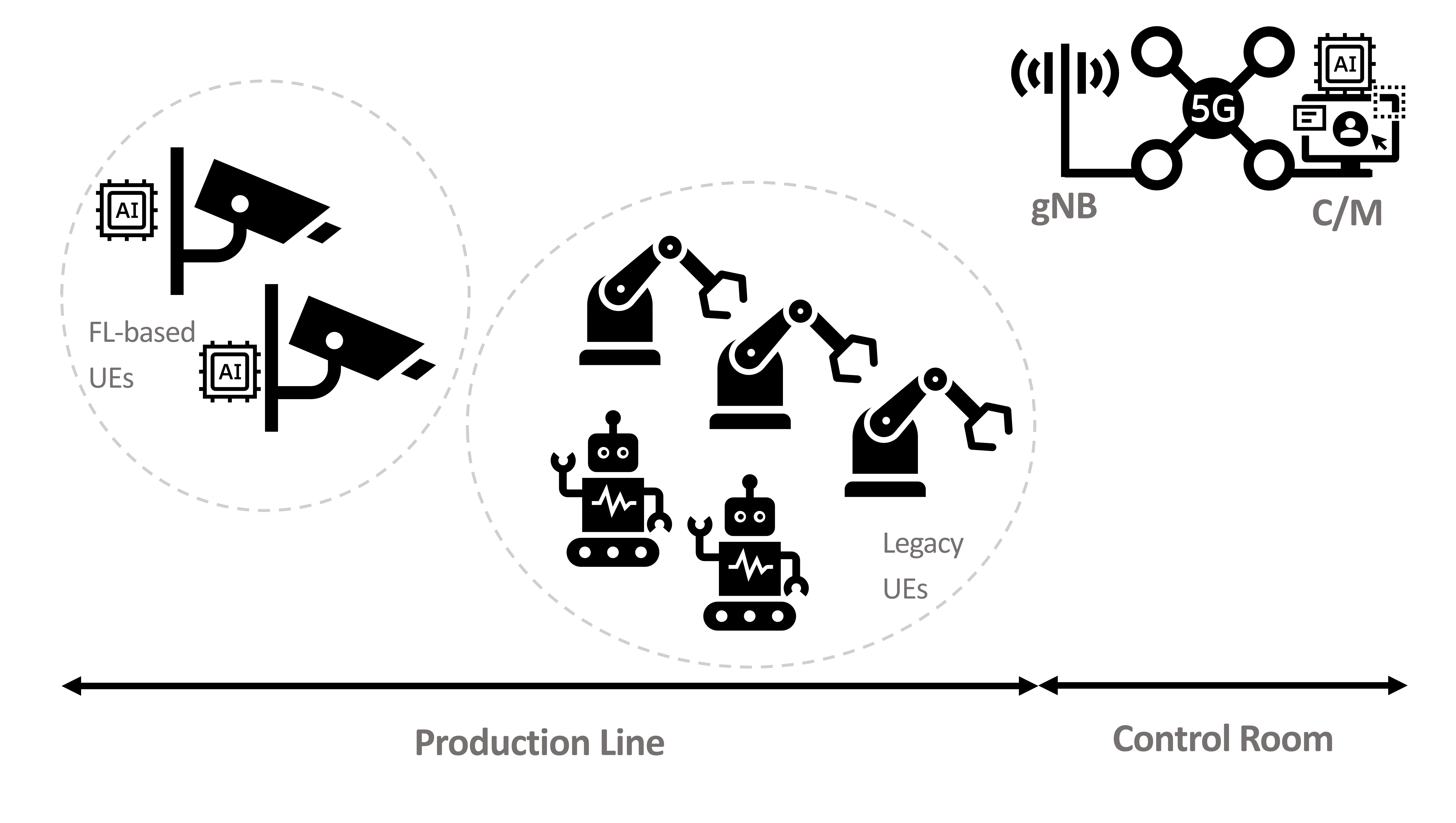}{The considered 2030-like industrial scenario, where industrial assets are equipped with legacy or \gls{fl}-based \glspl{ue} that are served by a private \gls{5g} network. These \glspl{ue} communicate with a \gls{cm} that is physically located in the control room of the factory.\label{fig:contention_scenario}}

This paper considers the 2030-like industrial scenario foreseen in \cite{hexax2021}, where a heterogeneous set of \gls{iiot} \glspl{ue} coexist in the same factory. In particular, Fig.~\ref{fig:contention_scenario} illustrates the considered network architecture, where $N$ industrial assets can either be legacy (e.g., robotic arms) or \gls{fl}-based (e.g., cameras performing image recognition through neural networks trained via \gls{fl}). Both types of entities are deployed in the production line of the factory and they have to communicate with a \gls{cm} located in the control room. To this aim, the industrial devices are equipped with \gls{5g} \glspl{ue} (one per industrial asset), and the factory is controlled by a private \gls{5g} network which consists of a dedicated \gls{ran} and \gls{5gc}. 

%%%%%%%%%%%%%%%%%%%%%%%%%%%%%%%%%%%%%%%%%%%%%%
\subsection{Deployment Model}
\Figure[t!](topskip=0pt, botskip=0pt, midskip=0pt)[width=3.25in]{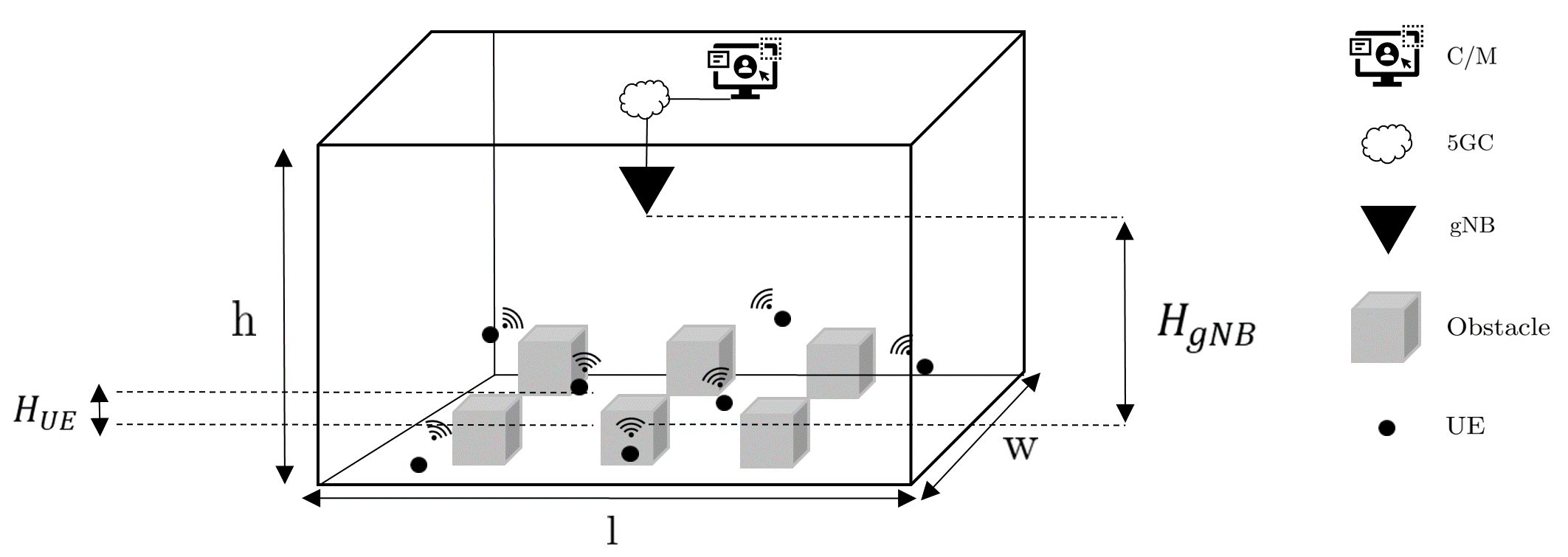}{The deployment model of the considered 2030-like industrial scenario.\label{fig:baseline_network_architecture}}

The factory floor has been modeled as a parallelepiped of length $l$, width $w$, and height $h$, as indicated in~\cite{5gaciaarchitecture}. Inside the factory, \gls{5g} communication between the \glspl{ue} and the \gls{gnb} can undergo severe attenuation due to obstructing elements (also referred to as \virg{obstacles} in the rest of the paper), such as walls or metal slabs. Obstacles are modeled as cubes and they are distributed inside the factory based on a given density $B_{\rm D}$, whereas $N$ \glspl{ue} are randomly and uniformly distributed inside the factory at a given height $H_{\rm UE}$ from the ground floor. The \gls{gnb} is instead located at height $H_{\rm gNB}$, as shown in Fig.~\ref{fig:baseline_network_architecture}.

%%%%%%%%%%%%%%%%%%%%%%%%%%%%%%%%%%%%%%%%%%%%%%
\subsection{Traffic Model}
\label{sec:traffic_model}

As anticipated in Sec.~\ref{sec:scenario}, we consider a factory containing two types of industrial assets. This means that we consider two different cathegories of \glspl{ue}, i.e., (i) \glspl{ue} that produce \gls{urllc} traffic (hereinafter referred to as \gls{urllc} \glspl{ue}) and (ii) \glspl{ue} that generate \gls{fl} traffic (hereinafter referred to as \gls{fl} \glspl{ue}). Hence, among the $N$ \glspl{ue} which are randomly and uniformly distributed in the factory, $N_{\rm UR}$ are \gls{urllc} \glspl{ue} and $N_{\rm FL}$ are \gls{fl} \glspl{ue}. 

%\textbf{Remark:} In real world implementations, \gls{urllc} and \gls{fl} traffics may also be generated by the same \gls{ue}, whereas here it is assumed that a \gls{ue} produces just one of the two. 

%%%%%%%%%%%%%%%%%%%%%%%%%%%%%%%%%%%%%%%%%%%%%%
\subsubsection{URLLC traffic}
It is modelled as a periodic bidirectional traffic, where \glspl{ue} transmit and receive application layer \glspl{pdu} of $P^{\rm U}_{\rm UR}$ and $P^{\rm D}_{\rm UR}$ bytes, respectively, with a fixed periodicity $\tau$. Depending on the transmission direction, that is, uplink or downlink, an application layer \gls{pdu} is discarded if it has been received with a delay exceeding $\tau_{\rm B}^{\rm U}$ or $\tau_{\rm B}^{\rm D}$, respectively. In particular, the delay is defined as the time elapsing from the instant when new data for transmission is generated at the sender, and the instant when it is entirely received by the recipient.  

%%%%%%%%%%%%%%%%%%%%%%%%%%%%%%%%%%%%%%%%%%%%%%
\subsubsection{FL traffic}
\label{sec:ai_traffic}

\Figure[t!](topskip=0pt, botskip=0pt, midskip=0pt)[width=3.25in]{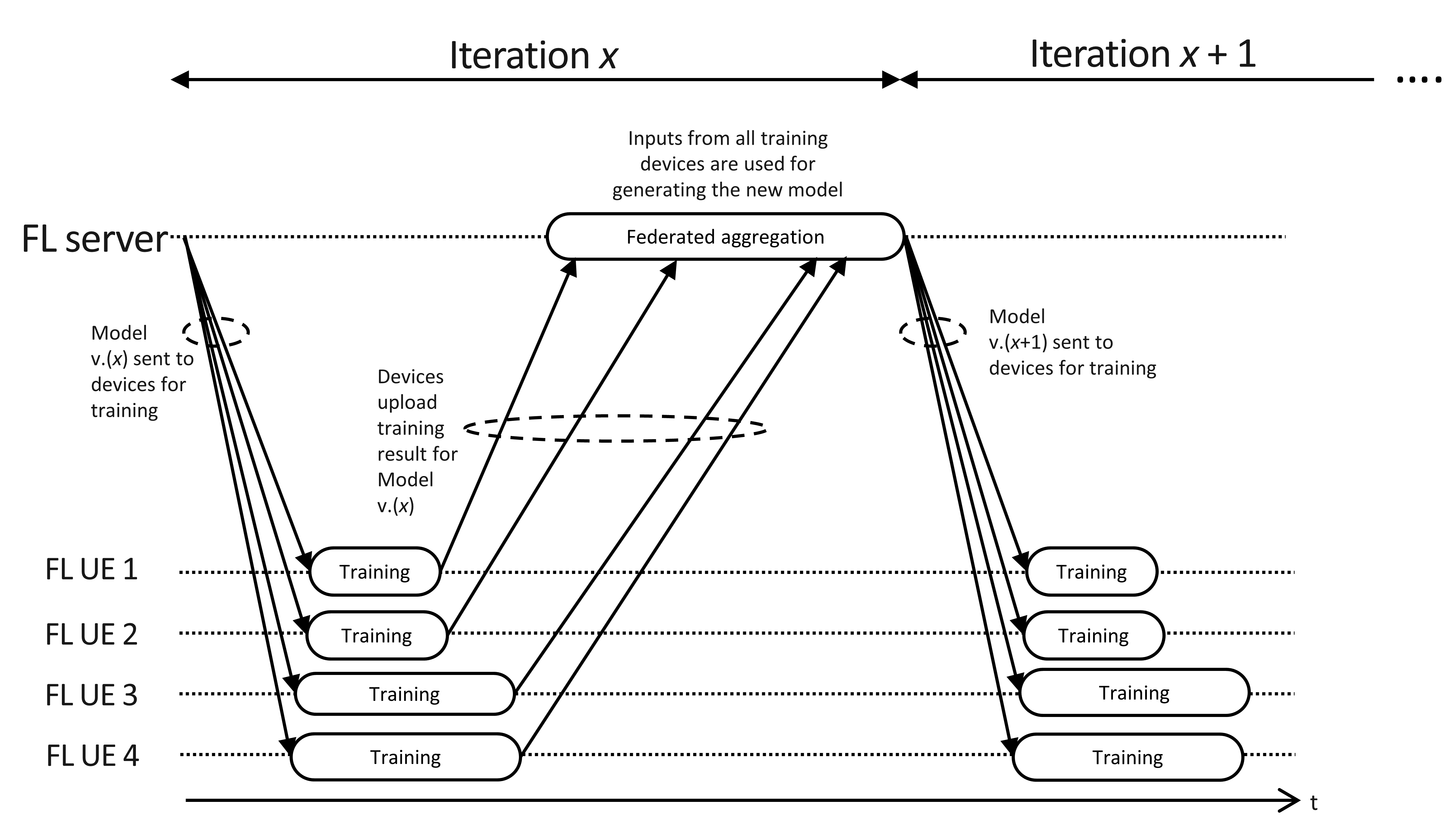}{Timing diagram of the considered \gls{fl} traffic, where a \gls{fl} server performs aggregation during iteration $x$ by using inputs from all \gls{fl} devices (four \glspl{ue} are depicted as an example), and it generates a new model version which will be sent during iteration $x$+1. For the sake of simplicity, the \gls{gnb} time axis is not shown, but it clearly acts as the forwarding node between \glspl{ue} and the \gls{fl} server.\label{fig:ai_traffic}}

It is modeled according to the synchronous \gls{fl} framework described in \cite{3gpp22874}, i.e., an iterative procedure where a \gls{fl} server, embedded in the \gls{cm}, trains a global model (e.g., the parameters of a neural network) by aggregating local models coming from the \gls{fl} devices. The approach of one, generic, \gls{fl} iteration is shown in Fig.~\ref{fig:ai_traffic}. At the beginning of the iteration, the \gls{fl} server sends the current version of the global model to the \gls{fl} \glspl{ue}, i.e., model v.($x$) during iteration $x$, where $x \in \{1,2, \dots, X\}$, being $X$ the total number of \gls{fl} iterations. Upon reception of the model, the devices perform local training and then send their updated version to the server. Finally, the server computes the new version of the model, i.e., model v.($x$+1), that will be sent in downlink during iteration $x$+1. Specifically, a unicast download of the model is assumed, that is, the server individually sends the same current version of the model to all \glspl{ue}\footnote{This choice avoids considering the technical difficulties of multicast transmissions, such as the selection of the \gls{mcs} \cite{araniti2017multicasting, condoluci2016enabling}.}. In this regard, $\tau^{\rm M}$ represents the time taken by the \gls{cm} application layer to send a \gls{fl} model towards the underlying transport protocol. From the communication perspective, having a unicast download means that \glspl{ue} will receive the model at different instants, and this spreads the subsequent uplink traffic over time, even due to potentially different training times of separate \glspl{ue}. However, in this synchronous \gls{fl} paradigm, the server has to receive all models before generating the new version. To avoid that the server stops due to errors (e.g., missing data fragments), we leverage \gls{tcp} at the transport layer, thus introducing retransmissions of lost data at layer 4 of the protocol stack. 

%%%%%%%%%%%%%%%%%%%%%%%%%%%%%%%%%%%%%%%%%%%%%%
\subsection{Channel Model}
The channel model is taken from \cite{3gpp38901}, where the blockage model B is used to determine the multipath attenuation caused by each of the obstacles using a knife-edge diffraction method, in addition to the path gain matrix and 3D channel data for all possible devices' locations.

%%%%%%%%%%%%%%%%%%%%%%%%%%%%%%%%%%%%%%%%%%%%%%
\subsection{Application of contention}
In such an \gls{iiot} scenario, the objective of this paper is to assess whether the \gls{cb} approach for \emph{uplink} transmissions (i.e., NR \gls{pusch}) described in Sec.~\ref{sec:contention_based_scheduling} can provide any benefit. In particular, we apply the \gls{cb} design for \gls{nr} \gls{pusch} to the \gls{fl} \glspl{ue} only, because there exist other scheduling algorithms, such as semi-persistent scheduling \cite{cuozzo2022enabling, arnjad2018latency, feng2019predictive}, which are better tailored to the \gls{urllc} traffic characteristics. For example, the stringent availability requirements of the \gls{urllc} traffic \cite{3gpp22804, etsi122104, 5gamericasurllc, ngmnreq}, cannot be easily met by a design where transmissions can also fail due to collisions in addition to channel impairments.

Conversely, the study of achievable performance with \gls{cb} strategies for the \gls{fl} traffic may be interesting due to the following reasons:

\begin{enumerate}
    \item It is likely spread over time due to (i) unicast download of the model, (ii) a non-negligible $\tau^{\rm M}$, and (iii) possibly different training times of the \glspl{ue}. Indeed, by design, \gls{cb} solutions work well when \glspl{ue} do not have to transmit at the same time;
    \item It is characterized by an on-off pattern, i.e., the download of a version of the model is followed by an upload of the new version and vice versa, but these two events never occur together. Since the download of a model is also characterized by the uplink transmissions of the corresponding \gls{tcp} \glspl{ack}, the immediate consequence of this property is that the uplink transmissions of \gls{tcp} \glspl{ack} and \gls{fl} models are not simultaneous and thus they cannot collide;
    \item \gls{fl} \glspl{ue} can, in principle, exploit their local \gls{ml} capabilities to also learn \emph{when} to use the \gls{cb} resources based on their past experience. However, this aspect is not considered in this study but it might be the subject of future works;
    %\item The \gls{fl} model sizes considered in this paper (see Sec.~\ref{sec:traffic_model}) are larger than the typical application layer \glspl{pdu} characterizing \gls{cb} uplink transmissions, since they rather involve transmissions of a few tens of bytes \cite{centenaro2017comparison, singh2017contention, au2014uplink}. %% DO NOT REMOVE %%
\end{enumerate}

%%%%%%%%%%%%%%%%%%%%%%%%%%%%%%%%%%%%%%%%%%%%%%
%%%%%%%%%%%%%%%%%%%%%%%%%%%%%%%%%%%%%%%%%%%%%%
% \section{Background on dynamic scheduling}
% Explain how dynamic scheduling works.

%%%%%%%%%%%%%%%%%%%%%%%%%%%%%%%%%%%%%%%%%%%%%%
%%%%%%%%%%%%%%%%%%%%%%%%%%%%%%%%%%%%%%%%%%%%%%
\section{Performance metrics}
\label{sec:performance_metrics}

This section describes the metrics used to assess the performance of the considered \gls{cb} design when applied to the \gls{fl} traffic and referring to the \gls{iiot} system model presented in Sec.~\ref{sec:system_model}. In particular, each metric refers to one traffic type, that is, either \gls{urllc} or \gls{fl} traffic.

%%%%%%%%%%%%%%%%%%%%%%%%%%%%%%%%%%%%%%%%%%%%%%%%%%%%%%%%%%%%%%%%%%%%%%%%%%%%%%
\subsection{Application layer availability}
Let us introduce a Bernoulli state variable for the $i$-th \gls{urllc} device, $X_i(t)$, that is zero if the last reception (at the application layer) has failed, either due to link failures or exceeding delay bound (see Sec.~\ref{sec:traffic_model}). Consequently, the application layer availability for the $i$-th \gls{urllc} \gls{ue} can be defined as follows:

\begin{equation}
a_i(t):= 
    \begin{cases}
    0, & \text{if $ \int_{t-T_{\rm SV}}^t X_i(\tau) \; d\tau = 0 $}  \\
    1, & \text{otherwise}
    \end{cases}
\end{equation}
where $T_{\rm SV}$ is the survival time, i.e., the interval of time during which the application can tolerate failures, i.e., missed reception of data. 

Therefore, the application layer availability for the $i$-th \gls{urllc} \gls{ue} can be written as:

\begin{equation}
    a_i := \lim_{T \to \infty} \frac{1}{T}  \int_{-\frac{T}{2}}^{\frac{T}{2}} a_i(t) \; dt 
\end{equation}

Finally, the average application layer availability, averaged over the total number of \gls{urllc} \glspl{ue} $N_{\rm UR}$, can be computed as $\overline{a} = \frac{\sum_{i=1}^{N_{\rm UR}} a_i}{N_{\rm UR}}$. Moreover, depending on the transmission direction (uplink or downlink), two average application layer availabilities can be defined, that is, $\overline{a^{\rm U}}$ and $\overline{a^{\rm D}}$. 

%%%%%%%%%%%%%%%%%%%%%%%%%%%%%%%%%%%%%%%%%%%%%%%%%%%%%%%%%%%%%%%%%%%%%%%%%%%%%%
\subsection{Collision probability}
The collision probability of the $n$-th \gls{fl} \gls{ue}, with $n \in \{ 0, 1, \dots, N_{\rm FL}\}$, is defined as follows:

\begin{equation}
    p^{\rm C}_{n} = \frac{C_{n}}{T_{n}}
\end{equation}

where $C_{n}$ is the number of \gls{cb} allocations where the $n$-th \gls{ue} has collided, and $T_{n}$ is the total number of utilized \gls{cb} resources. It immediately follows that the average collision probability, averaged over the total number of \gls{fl} \glspl{ue}, is $\overline{p^{\rm C}}=\frac{\sum_{n=1}^{N_{\rm FL}} p^{\rm C}_{n}}{N_{\rm FL}}$.

%%%%%%%%%%%%%%%%%%%%%%%%%%%%%%%%%%%%%%%%%%%%%%%%%%%%%%%%%%%%%%%%%%%%%%%%%%%%%%
\subsection{Model download time}

\Figure[t!](topskip=0pt, botskip=0pt, midskip=0pt)[width=3.25in]{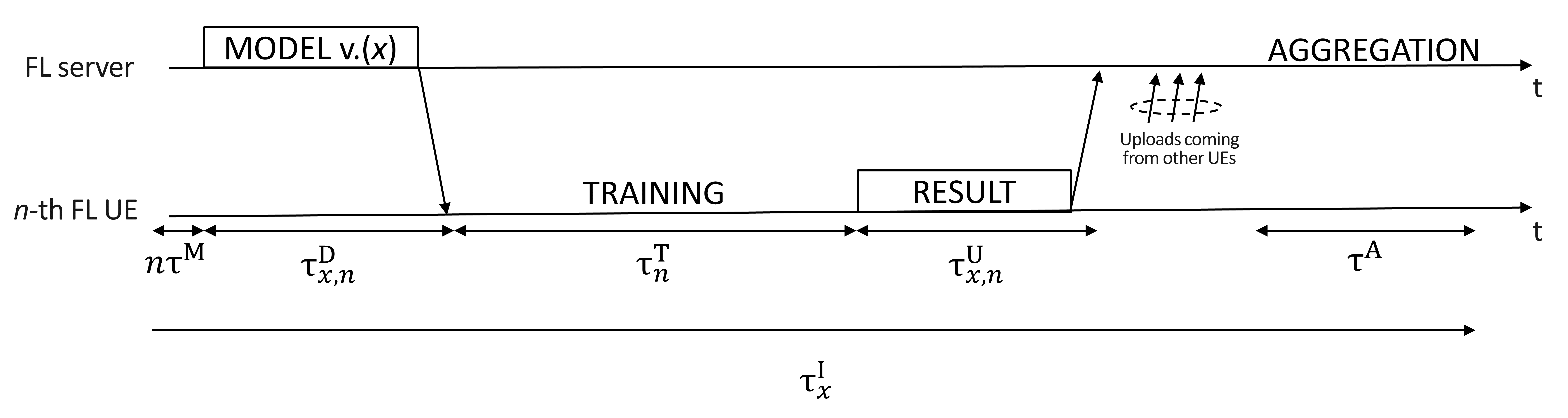}{Timing diagram of a generic \gls{fl} iteration $x$, where the download of model v.($x$) for the $n$-th \gls{fl} \gls{ue} starts after $n\tau^{\rm M}$ w.r.t the beginning of the iteration and lasts $\tau^{\rm D}_{x, n}$. When the download ends, the $n$-th \gls{fl} \gls{ue} performs training for $\tau^{\rm T}_n$ and then it takes $\tau^{\rm U}_{x, n}$ to upload the modified version of the model. Upon reception of the training outcomes from all \glspl{ue}, the \gls{fl} server ends iteration $x$ by taking $\tau^{\rm A}$ to create model v($x$+1).	\label{fig:ai_kpis}}

Fig.~\ref{fig:ai_kpis} formalizes the different timings characterizing a generic iteration $x$ and referring to the $n$-th \gls{fl} \gls{ue}. As already mentioned in Sec.~\ref{sec:traffic_model}, the iteration starts when the \gls{fl} server has a new model ready to be transmitted in unicast to all \gls{fl} \glspl{ue}, and the download of the model intended for the $n$-th \gls{fl} \gls{ue} starts after $n\tau^{\rm M}$ w.r.t the beginning of the iteration.

In this regard, the model download time $\tau^{\rm D}_{x, n}$ is defined as the time elapsing from the transmission of the first bit of model v.($x$) to the reception, by the $n$-th \gls{ue}, of its last bit. It immediately follows that the average model download time, averaged over the total number of \gls{fl} \glspl{ue} and iterations $X$, can be computed as $\overline{\tau^{\rm D}} = \frac{1}{X N_{\rm FL}}\sum_{x=1}^{X} \sum_{n=1}^{N_{\rm FL}} \tau^{\rm D}_{x, n}$.

%%%%%%%%%%%%%%%%%%%%%%%%%%%%%%%%%%%%%%%%%%%%%%%%%%%%%%%%%%%%%%%%%%%%%%%%%%%%%%
\subsection{Model upload time}
\label{sec:model_upload_time}
Upon receiving the model v.($x$), the $n$-th \gls{fl} \gls{ue} performs local training for a given amount of time $\tau^{\rm T}_n$. Afterward, it will transmit the result of the training, i.e., the updated version of the model, to the \gls{fl} server. 

Hence, the model upload time $\tau^{\rm U}_{x, n}$ (see Fig.~\ref{fig:ai_kpis}) is defined as the time elapsing from the transmission, by the $n$-th \gls{fl} \gls{ue}, of the first bit of the local updated version of model v.($x$), to the reception by the \gls{fl} server of its last bit. It immediately follows that the average model upload time, averaged over the total number of \gls{fl} \glspl{ue} and iterations $X$, can be computed as $\overline{\tau^{\rm U}} = \frac{1}{X N_{\rm FL}}\sum_{x=1}^{X} \sum_{n=1}^{N_{\rm FL}} \tau^{\rm U}_{x, n}$.  

%%%%%%%%%%%%%%%%%%%%%%%%%%%%%%%%%%%%%%%%%%%%%%%%%%%%%%%%%%%%%%%%%%%%%%%%%%%%%%
\subsection{Iteration time}
When the \gls{fl} server receives the updated versions of model v.($x$) from all the \gls{fl} \glspl{ue}, it takes $\tau^{\rm A}$ to perform aggregation, i.e., to generate the new version v.($x$+1). Since the iteration time $x$ is defined as the time elapsing from the generation of model v.($x$) to the creation of model v($x$+1) at the server-side, the average iteration time $\overline{\tau}$, averaged over the total number of iterations $X$, can be computed as $\overline{\tau^{\rm I}} = \frac{1}{X} \sum_{x=1}^{X} \tau^{\rm I}_x$.

%%%%%%%%%%%%%%%%%%%%%%%%%%%%%%%%%%%%%%%%%%%%%%
%%%%%%%%%%%%%%%%%%%%%%%%%%%%%%%%%%%%%%%%%%%%%%
\section{Performance Evaluation}
\label{sec:performance_evaluation}

%%%%%%%%%%%%%%%%%%%%%%%%%%%%%%%%%%%%%%%%%%%%%%
\subsection{Analyzed policies} 
This section briefly summarizes the different policies (and their nomenclature) used in the performance evaluation campaign.

%%%%%%%%%%%%%%%%%%%%%%%%%%%%%%%%%%%%%%%%%%%%%%
\subsubsection{Dynamic Scheduling}
It is the basic scheduling mechanism of \gls{5g} \gls{nr} described in Sec.~\ref{sec:uplink_data_transmissions_3gpp}, and it will be labeled as \textit{DS}.

%%%%%%%%%%%%%%%%%%%%%%%%%%%%%%%%%%%%%%%%%%%%%%
\subsubsection{Instantaneous Buffer Information}

\Figure[t!](topskip=0pt, botskip=0pt, midskip=0pt)[width=3.25in]{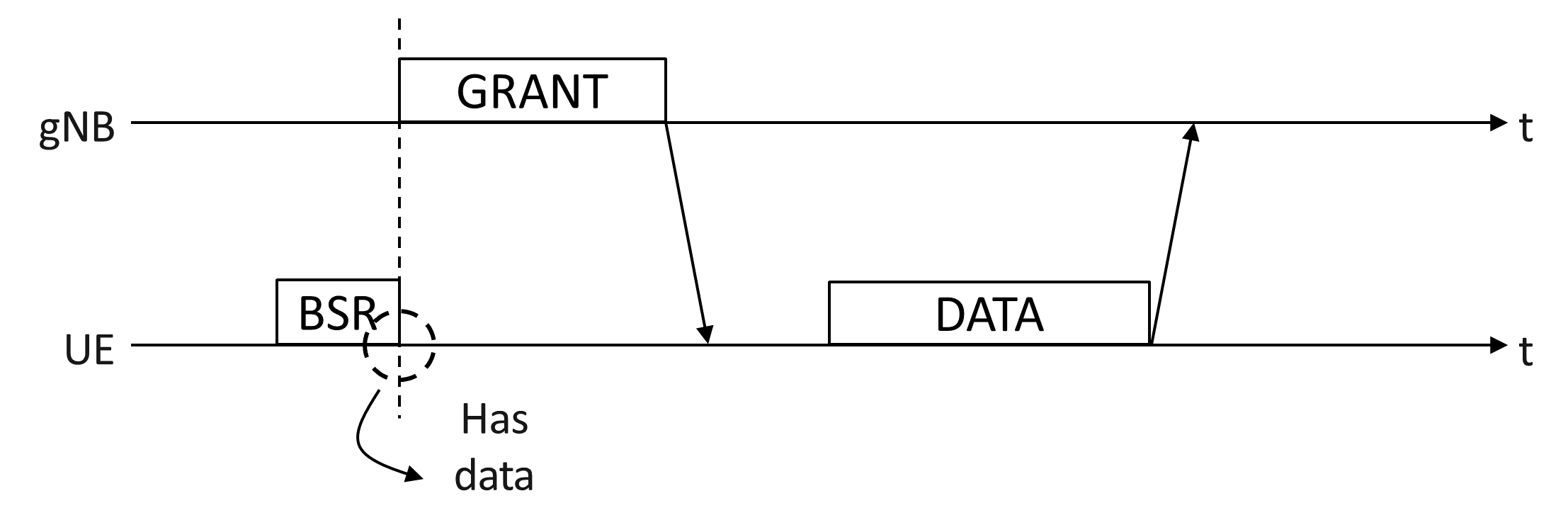}{Timing diagram of the \gls{ibi} approach, an ideal version of \gls{ds} used for comparison with \gls{cb} for \gls{nr} \gls{pusch}, where the \gls{gnb} immediately knows the \gls{bsr} of the \glspl{ue} as soon as they generate new data and it thus reserves dedicated radio resources for them.\label{fig:genie}}

It is an ideal version of \gls{ds} where the \gls{gnb} immediately knows the \gls{bsr} of the \glspl{ue} as soon as they generate new data, and it will be labeled as \textit{IBI}. More precisely, the \gls{ibi} approach is summarized in Fig.~\ref{fig:genie}, where, without any reception of \gls{sr} or \gls{bsr}, the \gls{gnb} reserves dedicated radio resources to the \glspl{ue} that have new data to transmit. %If there are enough resources available, the \gls{ue} can then transmit all the content of its queue; otherwise, the \gls{us} will reserve a new grant when it will be possible\footnote{\gls{fdd} is assumed, that is why, in Fig..~\ref{fig:genie}, downlink and uplink transmissions can happen simultaneously.} (this depends on the cell load, traffic priority, channel conditions, etc.). 
%It is worth highlighting that, in this scheme, the \gls{bsr} is not part of the \gls{mac} header, and thus it is not physically transmitted by the \gls{ue}, because the ideal assumption is that the \gls{gnb} immediately knows the \gls{bsr} at each new generation of data.

This kind of ideal version of \gls{ds} is useful for comparison with the considered \gls{cb} for \gls{nr} \gls{pusch} design because, in both cases, the \gls{sr} is not needed, i.e., the impact of the control plane is lower. However, differently from contention, no collisions are present in this case. Indeed, as it will be shown in Sec.~\ref{sec:numerical_results}, the considered \gls{cb} for \gls{nr} \gls{pusch} outperforms the \gls{ibi} scheme only in very specific occasions, and overall the performance of \gls{ibi} is close to the best achievable.

%%%%%%%%%%%%%%%%%%%%%%%%%%%%%%%%%%%%%%%%%%%%%%
\subsubsection{CB for NR PUSCH with retransmissions on dedicated resources}
It refers to the \gls{cb} design for \gls{nr} \gls{pusch} described in Sec.~\ref{sec:retx_mechanisms}, where retransmissions are scheduled via dedicated resources, and it will be labeled as \textit{\gls{cb} for \gls{nr} \gls{pusch} with re-tx on dedicated}.

%%%%%%%%%%%%%%%%%%%%%%%%%%%%%%%%%%%%%%%%%%%%%%
\subsubsection{CB for NR PUSCH with retransmissions on contention resources}
It refers to the \gls{cb} design for \gls{nr} \gls{pusch} described in Sec.~\ref{sec:retx_mechanisms}, where retransmissions are scheduled via \gls{cb} resources, and it will be labeled as \textit{\gls{cb} for \gls{nr} \gls{pusch} with re-tx on contention}.

%%%%%%%%%%%%%%%%%%%%%%%%%%%%%%%%%%%%%%%%%%%%%%
\subsection{Numerical results} 
\label{sec:numerical_results}

\def\arraystretch{1.2}
\begin{table}[!t]
\centering
\footnotesize
\caption{Simulation parameters.}
\label{tab:system_parameter_settings}
\begin{tabular}{|c|c|c|}
\hline
\textbf{Parameter} & \textbf{Description} & \textbf{Value}  \\\hline
$f_c$ & Carrier frequency & 2.6 GHz\\ \hline
$B$ & Overall system bandwidth & 40 MHz \\ \hline
$\Delta f$ & Subcarrier spacing & 30 kHz\\ \hline
$T_{\rm S}$ & Simulation time & 150 s\\  \hline
$A_{\rm gNB}$ &  \gls{gnb} antenna number & 2 \\ \hline  
$A_{\rm UE}$ & \glspl{ue} antenna number  & 1 \\ \hline   
$P_{\rm TX}^{\rm U}$ & \gls{ue} transmit power & 0.2 W \\ \hline  
$P_{\rm TX}^{\rm d}$& \gls{gnb} transmit power & 0.5 W \\ \hline 
$B_{\rm D}$ & Obstacle's density & 0.15 obstacles/$m^2$ \\   \hline
$S$ & Side of the obstacles & 9 m \\  \hline
$l$ & Length of the factory floor & 15 m~\cite{5gaciaarchitecture}\\   \hline
$w$ & Width of the factory floor & 15 m\\  \hline
$h$ & Height of the factory floor & 11 m \\  \hline
$H_{\rm UE}$ & \glspl{ue}' height & 1.5 m \\ \hline
$H_{\rm gNB}$ & \gls{gnb} height & 10 m \\ \hline
$N_{\rm UR}$ & Number of \gls{urllc} \glspl{ue} & 10 \\ \hline
$P^{\rm U}_{\rm UR}$ & Uplink application layer \gls{pdu} & 64 B \\ & for \gls{urllc} \glspl{ue} &  \\   \hline
$P^{\rm D}_{\rm UR}$ & Downlink application layer \gls{pdu} & 80 B \\  & for \gls{urllc} \glspl{ue} &  \\  \hline
$\tau_{\rm B}^{\rm U}$ & \gls{urllc} uplink delay bound & 10 ms \\   \hline
$\tau_{\rm B}^{\rm D}$ & \gls{urllc} downlink delay bound & 3 ms \\   \hline
$\tau$ & \gls{urllc} uplink/downlink & 5 ms \\ & transmission periodicity &  \\   \hline
%$P_{\rm FL}$ & Application layer \gls{pdu} & \{2, 5.2, \\  & for \gls{fl} \glspl{ue} & 10.4, 16.8\} MB  \\  \hline
$\tau^{\rm M}$ & Time taken to transfer the \gls{fl} model & 10 ms \\ & to the underlying layers & \\   \hline
$\tau^{\rm T}$ & Training time of \gls{fl} \glspl{ue} & 10 s \\   \hline
$\tau^{\rm A}$ & Aggregation time of the \gls{fl} server & 10 s \\  \hline
$T_{\rm BO}$ & Backoff interval & 10 slots \\ \hline
$T_{\rm SV}$ & Survival time & 15 ms \\ \hline
%$N_{\rm RX}$ & Maximum number of \gls{harq} & \{0, 10\} \\ & uplink/downlink retransmissions &  \\ \hline 
\end{tabular}
\vspace{-3ex}
\end{table}

Simulation parameters, if not otherwise specified, are reported in Table~\ref{tab:system_parameter_settings}. In particular, some additional information should be provided:

\begin{itemize}
    \item Based on the chosen numerology (i.e., $\Delta f = 30$ kHz), the \gls{5g} slots are 0.5 ms long. Hence, each simulation is formed by 300000 slots since $T_{\rm S} = 150$ seconds. All results have been obtained by averaging over 10 simulations, that is 3000000 slots, where each simulation was associated with a diverse seed, i.e., a different distribution of \glspl{ue} and obstacles, as well as an independent channel evolution;
    \item The total number of \gls{fl} iterations, $X$, is not fixed a priori (and thus it does not appear in Table~\ref{tab:system_parameter_settings}) since it depends on the specific settings of a given simulation run, such as the number of \glspl{ue}, the model size, the considered retransmission policy, etc.;
    \item Based on the chosen bandwidth and numerology, the overall number of \glspl{rb} is 112, and the latter constitutes the upper bound for the size of the \gls{cb} grant in the frequency domain. Conversely, there is no limit in the time domain, i.e., on the available \gls{ofdm} symbols to be used for \gls{cb} uplink transmissions; 
    \item As far as the \gls{fl} model is concerned, we follow the approach of \cite{3gpp22874}, where both the \gls{fl} server and \glspl{ue} are assumed to implement a \gls{dnn} following the MobileNets architecture \cite{howard2017mobilenets}, i.e., a class of \glspl{dnn} models based on a streamlined architecture for mobile and embedded vision applications. However, it is important to underline that we did not implement any \gls{dnn} because we focus here on the communication part of this traffic. Hence, we set the related \gls{fl} timings (i.e., training and aggregation time) and model sizes based on the study in \cite{howard2017mobilenets}; %Specifically, the MobileNet-224 architecture can have $\{0.5, 1.3, 2.6, 4.2\}$ million parameters; by considering 32 bits for each parameter, the application layer \gls{pdu} of these models (denoted as $P_{\rm FL}$ in the following) is thus represented by $\{2, 5.2, 10.4, 16.8\}$ MB, both in uplink and downlink;
    \item All results show a confidence interval with a probability of 95\%;
    % \item In all simulations, 10 \gls{urllc} \glspl{ue} are scheduled via \gls{ds} with an \gls{sr} periodicity of 10 slots, that is, 5 ms;
    \item When \gls{fl} \glspl{ue} have to rely on \gls{ds}, \glspl{sr} periodicity is 1 slot, i.e., 0.5 ms;    
    \item As already described in Sec.~\ref{sec:contention_based_scheduling}, the maximum number of \gls{harq} uplink/downlink retransmissions, $N_{\rm RX}$, is set to 0 when \gls{fl} \glspl{ue} retransmit via contention, otherwise it is set to 10. On the other hand, $N_{\rm RX}$ is set to 3 or 2 for uplink and downlink \gls{urllc} transmissions, respectively;
    %\item The maximum number of \gls{rlc} uplink/downlink retransmissions is 8. When this number is reached, \gls{rlc} declares a \gls{rlf}, i.e., the user has to re-connect to the network;
    \item Among the same category of \glspl{ue}, proportional fair is used as the radio resource assignment algorithm \cite{goldsmith2005wireless};
    \item \gls{urllc} \glspl{ue} have higher priority w.r.t \gls{fl} \glspl{ue}. Remarkably, retransmissions have a higher priority w.r.t first transmissions, and this means that retransmissions of \gls{fl} \gls{tb} are prioritized w.r.t first transmissions of \gls{urllc} \gls{tb}. In a nutshell, the different cases can be sorted in descending priority order as follows:
    
    \begin{enumerate}
        \item Retransmissions of \gls{urllc} \glspl{ue};
        \item Dedicated retransmissions of \gls{fl} \glspl{ue} (when considering \gls{cb} for \gls{nr} \gls{pusch} with re-tx on dedicated, as described in Sec.~\ref{sec:re-tx_dedicated});
        \item First transmissions of \gls{urllc} \glspl{ue};
        \item First transmissions or \gls{cb} retransmissions of \gls{fl} \glspl{ue} (when considering \gls{cb} for \gls{nr} \gls{pusch} with re-tx on contention, as described in Sec.~\ref{sec:re-tx_contention}).
    \end{enumerate}
    
\end{itemize}

\begin{figure*}[!t]
\centering
	\subfloat[Model upload time as a function of the number of \gls{fl} \glspl{ue} when $P_{\rm FL}=12$ kB.]
	{\includegraphics[width=0.4\linewidth]{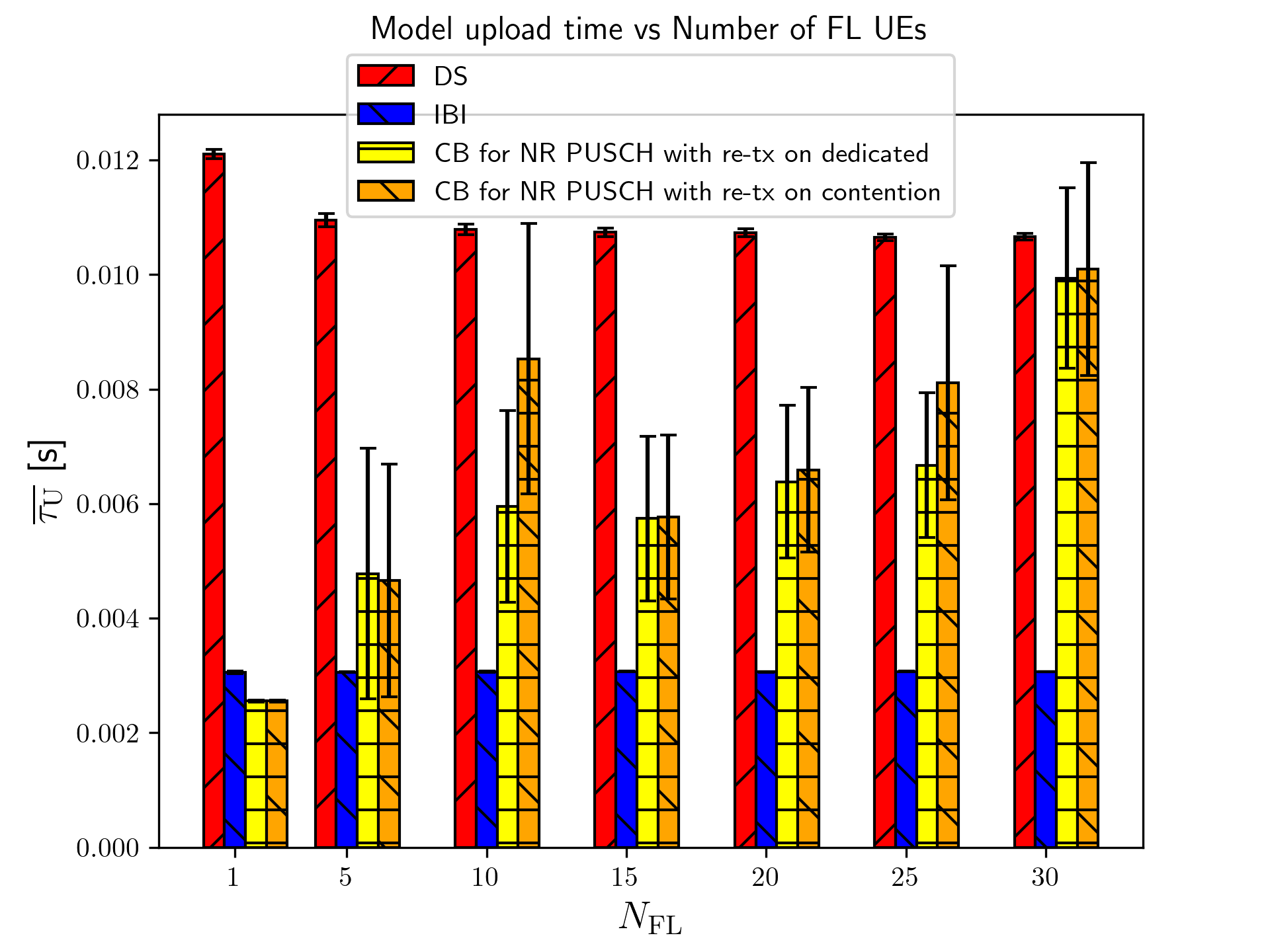}
	\label{fig:upload_time_12kb}}
    \subfloat[Model upload time as a function of the number of \gls{fl} \glspl{ue} when $P_{\rm FL}=2$ MB.]
	{\includegraphics[width=0.4\linewidth]{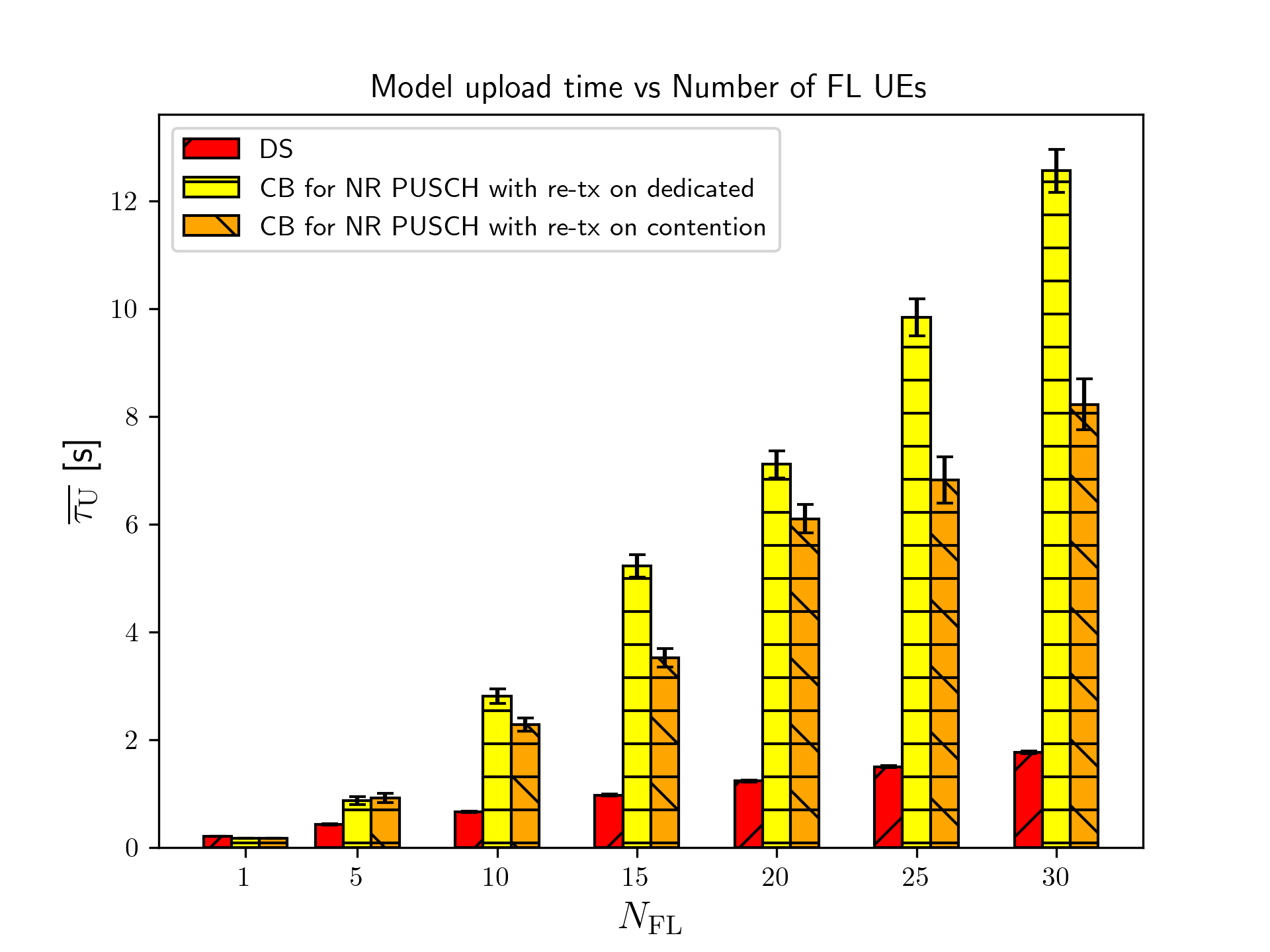}
	\label{fig:upload_time_2MB}} \\
	\subfloat[Model download time as a function of the number of \gls{fl} \glspl{ue} when $P_{\rm FL}=16$ kB.]
	{\includegraphics[width=0.4\linewidth]{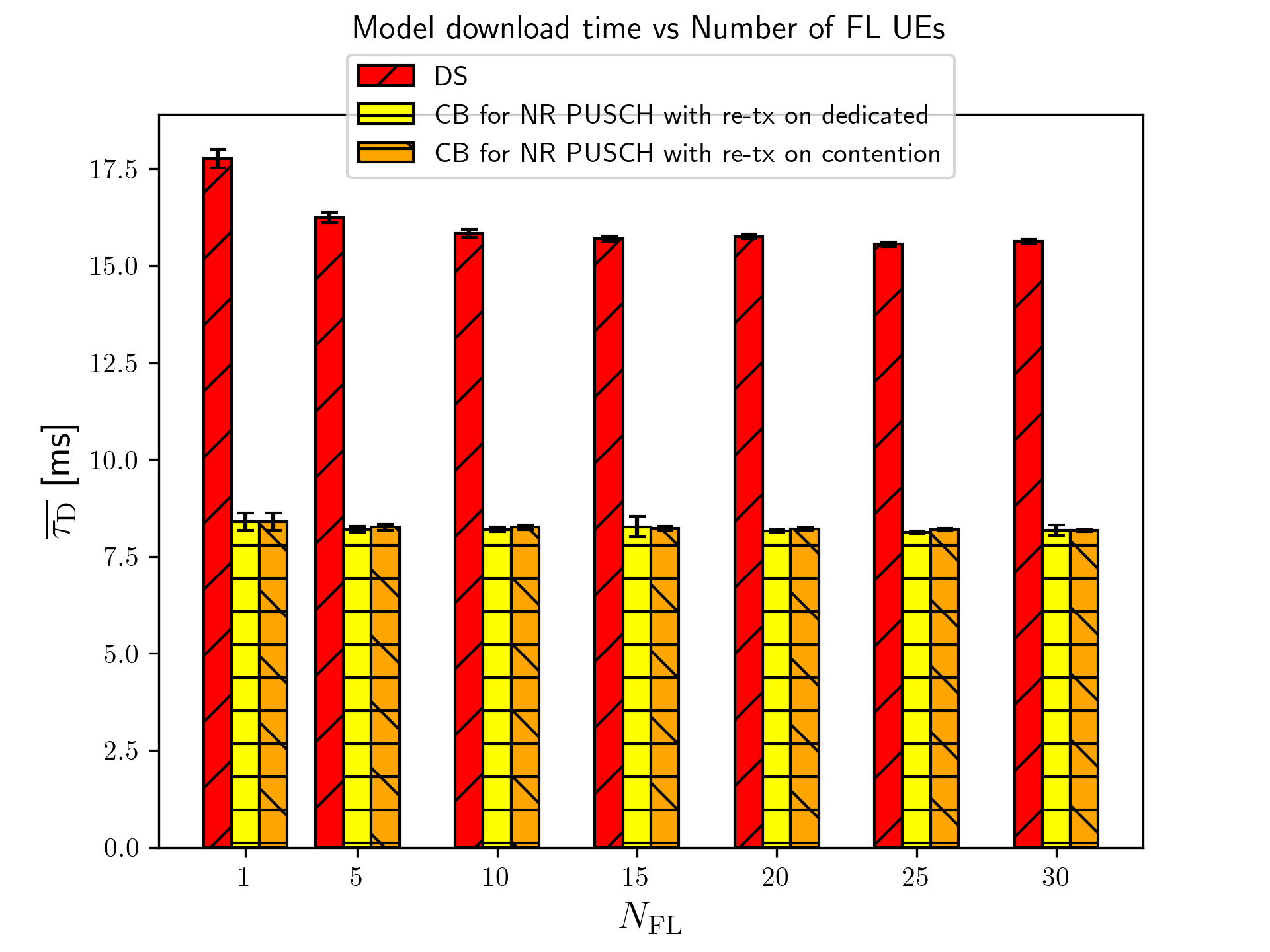}
	\label{fig:download_time_16kB}}
	\subfloat[Model download time as a function of the number of \gls{fl} \glspl{ue} when $P_{\rm FL}=2$ MB. ]{\includegraphics[width=0.4\linewidth]{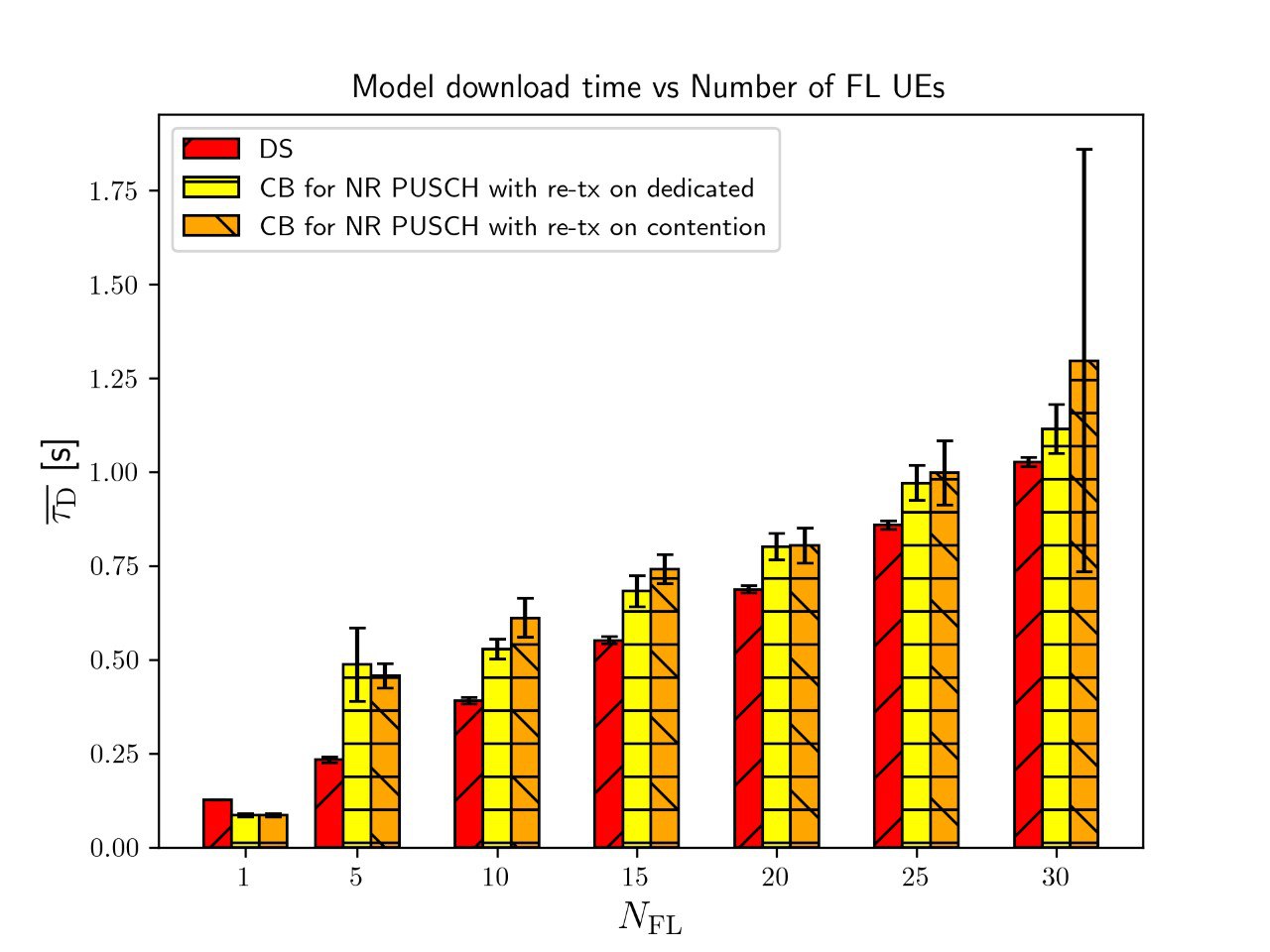}
    \label{fig:download_time_2MB}}
\caption{Model upload/download time as a function of the number of \gls{fl} \glspl{ue}, and by considering model sizes of 12/16 kB and 2 MB. The different curves represent four different situations, that is, the \gls{fl} \glspl{ue} are scheduled (i) via \gls{ds}, or (ii) \gls{ibi}, or (iii) perform \gls{cb} for \gls{nr} \gls{pusch} with re-tx on dedicated resources, (iv) or perform \gls{cb} for \gls{nr} \gls{pusch} with re-tx on \gls{cb} resources.}
\label{fig:transfer_times}
\end{figure*}

\begin{figure*}[!t]
	\subfloat[Case with $P_{\rm FL}=12$ kB.]
	{\includegraphics[width=0.5\linewidth]{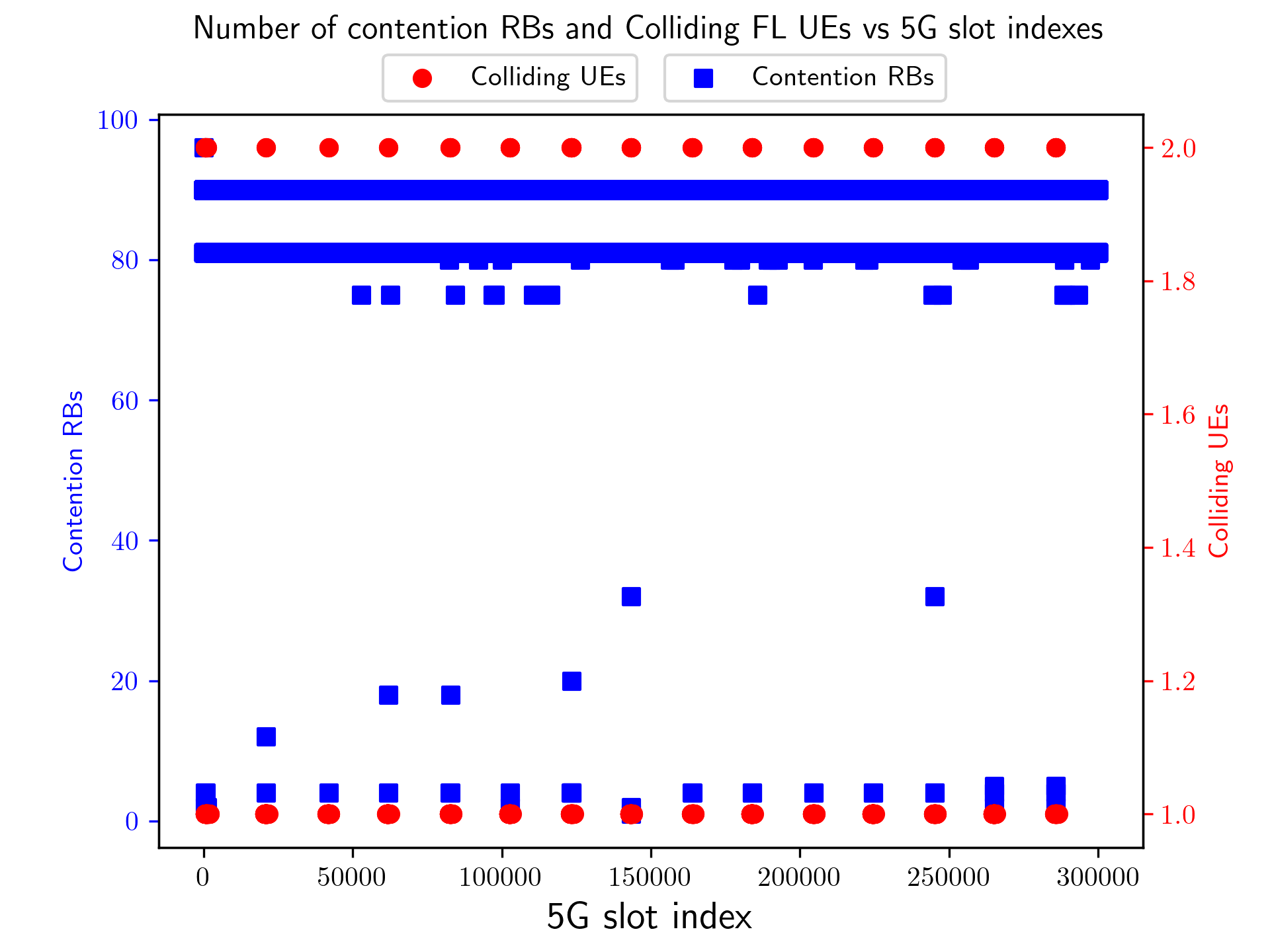}
	\label{fig:window_closure_12kB}}
    \subfloat[Case with $P_{\rm FL}=2$ MB.]
    {\includegraphics[width=0.5\linewidth]{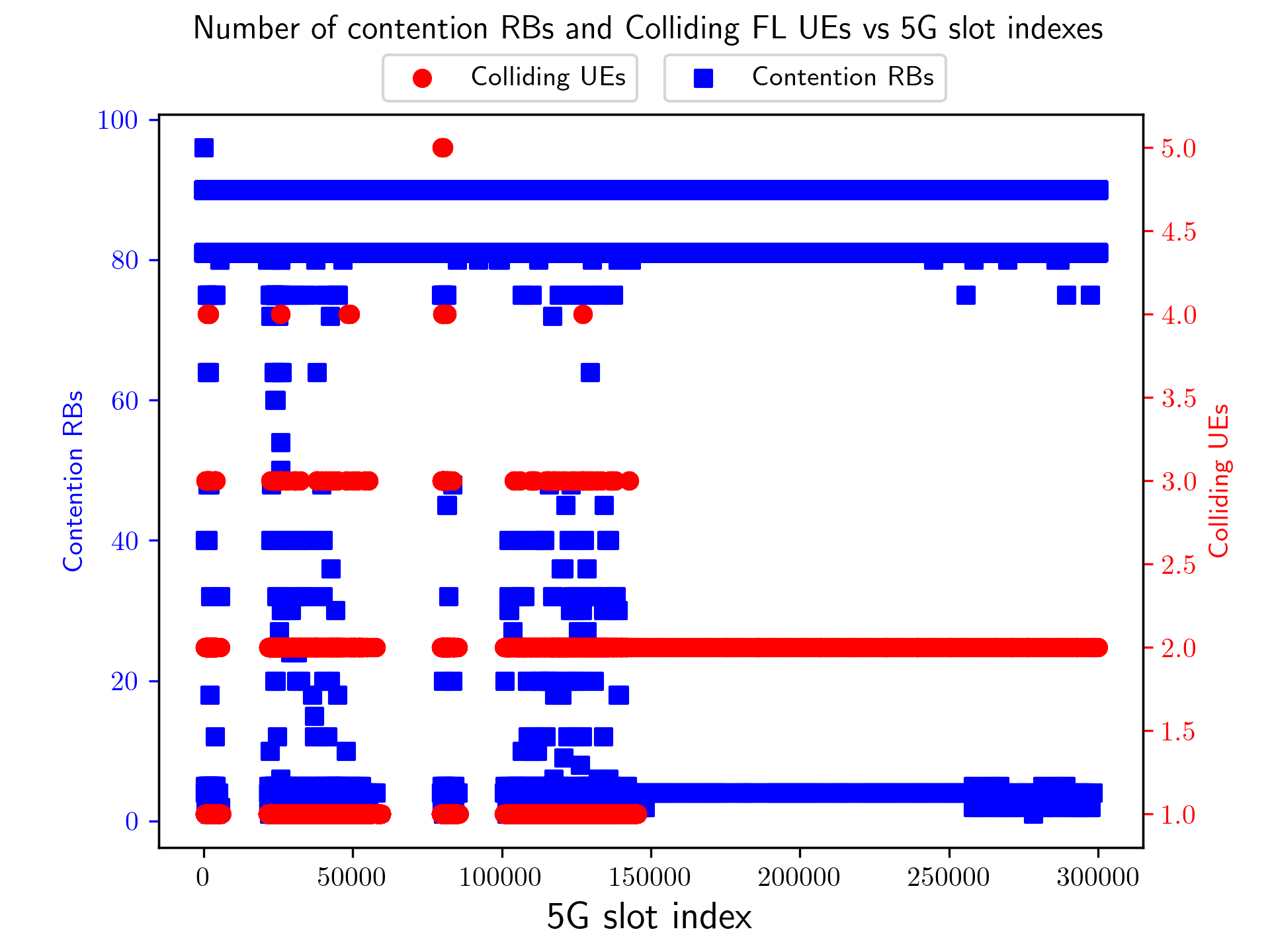}
	\label{fig:contention_window_closure_2MB}}
\caption{Number of \glspl{rb} used for \gls{cb} allocations (blue dots) and number of colliding \gls{fl} \glspl{ue} (red dots), as a function of the \gls{5g} slot indexes contained in one simulation run, when considering 30 \gls{fl} \glspl{ue} transmitting and receiving models made of 12 kB (on the left) or 2 MB (on the right) by means of \gls{cb} for NR \gls{pusch} with retransmissions on dedicated resources.}
\label{fig:window_closure}
\end{figure*}

All that being said, Fig.~\ref{fig:transfer_times} is a collection of four plots showing the average model upload/download time as a function of the number of \gls{fl} \glspl{ue}, and by considering \gls{fl} model sizes, $P_{\rm FL}$, of 12 and 16 kB, as well as 2 MB (the latter is the minimum size considered in \cite{howard2017mobilenets}). %In particular, the \gls{fl} \glspl{ue} can be served with dynamic scheduled (labeled as \textit{ds}), or use \gls{cb} for the first transmission and \gls{ds} for retransmissions (labeled as \textit{CB + re-tx on DS}), or use \gls{cb} for both first transmission or retransmissions (labeled as \textit{CB + re-tx on CB}), or be dynamically scheduled through the \gls{ibi} (labeled as \textit{IBI}). 
It is clearly evident that the considered \gls{cb} approach outperforms \gls{ds} when the model size is 12 and 16 kB, for all the considered values of $N_{\rm FL}$, both in upload and download. Notice that the considered \gls{cb} for NR \gls{pusch} is applied to any uplink communication of the \gls{fl} \glspl{ue}; therefore, the gain in the model download time is due to an improvement of the time needed to transmit the \gls{tcp} \glspl{ack}. \newline 
However, the gain of \gls{cb} for \gls{nr} \gls{pusch} ceases to be true for a larger model size of 2 MB, even for a small number of \gls{fl} \glspl{ue} (i.e., $N_{\rm FL}>$ 1), and this is more evident for the model upload time. Indeed, in upload, a larger model size increases both the number of transmissions (due to segmentation \cite{parkvall20185g}) and their dimensions, whereas, in download, the number of \gls{tcp} \glspl{ack} transmissions increases but their size remains unaffected. \newline 
The two retransmission policies perform similarly, especially for low model sizes, because of a sufficiently low amount of new collisions during the retransmissions. Nonetheless, counterintuitively, Fig.~\ref{fig:upload_time_2MB} shows lower model upload times when retransmitting via \gls{cb} grants. The reason for that is two-fold. On the one hand, reserving dedicated resources for retransmissions of a non-negligible size significantly shrinks the available resources that can be used for future \gls{cb} allocations (due to the higher priority given to retransmissions w.r.t first transmissions), thereby prolonging the overall transfer time of the case with retransmissions on dedicated resources (this aspect will be better clarified later when discussing Fig.~\ref{fig:window_closure}). On the other hand, looking at behaviors of single \glspl{ue}, we noticed that the number of \glspl{ue} which do \emph{not} complete all iterations within the simulation time is higher (on average) when retransmissions happen via \gls{cb} resources compared to retransmitting on dedicated resources (the reader can recall from Fig.~\ref{fig:ai_traffic} that a \gls{ue} finalizes an iteration only when it concludes its model upload). Due to this second reason, the lower model upload time for the case of retransmissions on \gls{cb} resources when $P_{\rm FL}=2$ MB is not fully reflecting a better performance from a \gls{fl} iteration point of view compared to retransmissions on dedicated resources. Indeed, the latter approach, although with higher upload transfer times, allows more \glspl{ue} to complete the \gls{fl} iterations also in case of higher load as retransmissions are (i) without collisions, and (ii) can exploit a per-\gls{ue} link adaptation process for the \gls{mcs} selection. \newline  
As expected, \gls{cb} for NR \gls{pusch} outperforms \gls{ibi} only in a very specific case, i.e., when considering a single \gls{fl} \gls{ue} for small model sizes. This is because, the absence of collisions highlights the gain provided by \gls{cb}, i.e., the single \gls{ue} is likely to immediately perform the few transmissions needed to upload the small model because it already received the \gls{cb} grant when the data is generated (see Fig.~\ref{fig:timing_diagram_optimistic}). %might likely have already some resources for data transmission (hence, no extra time spent waiting for the reception of a dedicated grant) and, with only one \gls{ue}, there are no collision events that the few transmissions needed to transfer the small model size are performed quickly. 
When the number of \glspl{ue} increases, \gls{ibi} remains the best-performing policy due to its ideality. For this reason, we will not further consider the performance of \gls{ibi} in the analysis. 

To explain the reasons behind the choice of 12 and 16 kB as the \gls{fl} model sizes for Figs.~\ref{fig:upload_time_12kb}  and \ref{fig:download_time_16kB}, Table~\ref{tab:model_upload_gain} shows the conditions under which \gls{cb} for \gls{nr} \gls{pusch} with retransmissions on dedicated resources provides lower model upload times w.r.t \gls{ds} as a function of the model size, $P_{\rm FL}$, and maximum bandwidth allowed for \gls{cb} transmissions $B_{\rm CB}$ (out of the overall system bandwidth $B$) when considering $N_{\rm FL}=$ 30. As can be seen, independently of the considered $B_{\rm CB}$ value, a model size of 12 kB is the maximum value for which the considered \gls{cb} design provides benefits over \gls{ds}, thereby motivating the model size choice of Fig.~\ref{fig:upload_time_12kb}. Of course, the same holds for the model download times, i.e., 16 kB is the maximum model size for which there are gains, as well as when considering the design employing retransmissions on \gls{cb} resources.

\begin{table}[!t]
\centering
\footnotesize
\caption{Conditions under which \gls{cb} for \gls{nr} \gls{pusch} with retransmissions on dedicated resources provides lower model upload times w.r.t \gls{ds} as a function of the model size $P_{\rm FL}$ and maximum bandwidth allowed for \gls{cb} transmissions $B_{\rm CB}$ when considering $N_{\rm FL}=$ 30.}
\label{tab:model_upload_gain}
\begin{tabular}{cc|c|c|c|c|}
\cline{3-6}
& & \multicolumn{4}{c|}{$P_{\rm FL}$ [kB]} \\
\cline{3-6}
{} & {} & $\mathbf{1}$ & $\mathbf{8}$ & $\mathbf{12}$ & $\mathbf{16}$  \\ \hline
\multicolumn{1}{|c|}{\multirow{4}{*}{$B_{\rm CB}$ [MHz]}} & $\mathbf{5}$ & YES & NO & NO & NO  \\ \cline{2-6}
\multicolumn{1}{|c|}{} & $\mathbf{10}$ & YES & YES & YES (up to 5 UEs) & NO  \\ \cline{2-6}
\multicolumn{1}{|c|}{} & $\mathbf{20}$ & YES & YES & YES (up to 20 UEs) & NO  \\ \cline{2-6}
\multicolumn{1}{|c|}{} & $\mathbf{40}$ & YES & YES & YES & NO  \\ \hline
\end{tabular}
\end{table}

\begin{table}[!t]
\centering
\footnotesize
\caption{Average collision probability and average iteration time as a function of the number of \gls{fl} \glspl{ue}, when considering the model size of 2 MB, and the considered \gls{cb} approach for NR \gls{pusch} with retransmissions on dedicated resources.}
\label{tab:collision_prob_iteration_times}
\begin{tabular}{|c|c|c|}
\hline
\textbf{$N_{\rm FL}$} & \textbf{$\overline{p^{\rm C}}$} & \textbf{$\overline{\tau^{\rm I}}$ [s]}  \\\hline
1 & 0 & 20.2498 \\ \hline 
5 & 0.0206 & 22.2051 \\ \hline
10 & 0.0426 & 25.1477 \\ \hline
15 & 0.0649 & 29.6163 \\  \hline
20 & 0.0826 &  32.1696 \\ \hline  
25 & 0.1017 &  37.2457 \\ \hline   
30 & 0.1238 & 42.4982 \\ \hline  
\end{tabular}
\end{table}

Next, in Table~\ref{tab:collision_prob_iteration_times} we show the average collision probability, $\overline{p^{\rm C}}$, and average iteration time, $\overline{\tau^{\rm I}}$, when considering the model size of 2 MB, and the considered \gls{cb} for NR \gls{pusch} with retransmissions on dedicated resources. It can be seen that the average iteration times can be tens of seconds even for low loads, thus indicating that a \gls{fl} training can be quite large if it involved higher loads (i.e., a non-negligible training phase has to be performed before having \gls{fl}-based cameras ready to perform image recognition). 

It is interesting to notice that the average collision probabilities are relatively small (at most $\sim$ 12\%). This consideration %, together with noting that, when considering high loads, retransmitting on \gls{cb} resources is preferable in upload (see Fig.~\ref{fig:upload_time_2MB}), whereas retransmitting on dedicated resources remain a viable option in download (see Fig.~\ref{fig:download_time_2MB}), 
suggests that the introduction of a collision framework in a \gls{5g} \gls{nr} \gls{iiot} network produces an additional phenomenon that cannot be merely controlled by looking only at the collision probability. This thought is confirmed through Fig.~\ref{fig:window_closure}, where it illustrates both, the number of \glspl{rb} used for \gls{cb} allocations (blue dots) and the number of colliding \gls{fl} \glspl{ue} (red dots), as a function of the \gls{5g} slot indexes contained in one simulation run. Two model sizes are compared, i.e., 12 kB (on the left) and 2 MB (on the right). The plot refers to one cell, and a total amount of 30 \gls{fl} \glspl{ue} which are scheduled via \gls{cb} for \gls{nr} \gls{pusch} with retransmissions on dedicated resources. Among the total amount of 112 \glspl{rb}, the \gls{gnb} never allocates more than 96 \glspl{rb} for the \gls{cb} allocation due to the presence of the higher priority always-on \gls{urllc} traffic. With a fixed periodicity, the \gls{cb} allocation shrinks to 81 \glspl{rb} due to periodic control plane signals, such as \glspl{cqi}. However, it can be clearly noted that, when considering model sizes of 12 kB, the number of collisions is sporadic and they never involve more than 2 \glspl{ue}. Consequently, the number of \glspl{rb} used for the \gls{cb} allocation (blue dots) remains high for most of the time. This ceases to be true when considering model sizes of 2 MB, because, for the vast majority of the simulation, the \gls{cb} allocation is shrinked to a few \glspl{rb} (close to 5), thus resulting in very long download/upload transfer times (as previously shown in Fig.~\ref{fig:upload_time_2MB} and \ref{fig:download_time_2MB}). Indeed, the size of the model is such that, even the retransmissions of a few number of colliding \glspl{ue} (no more than 5) need most of the dedicated radio resources. Consequently, the higher priority given to such retransmissions dramatically reduces the amount of resources that can be used for future first transmissions via \gls{cb} allocations. This phenomenon explains the performance degradation of the considered \gls{cb} scheme when dealing with larger model size despite a relatively low collision probability, and, at the same time, it motivates the benefit in retransmitting on \gls{cb} resources for this specific case (as explained when describing Fig.~\ref{fig:upload_time_2MB}).

\begin{figure*}[!t]
	\subfloat[Average uplink availability of 10 \gls{urllc} \glspl{ue}.]
	{\includegraphics[width=0.5\linewidth]{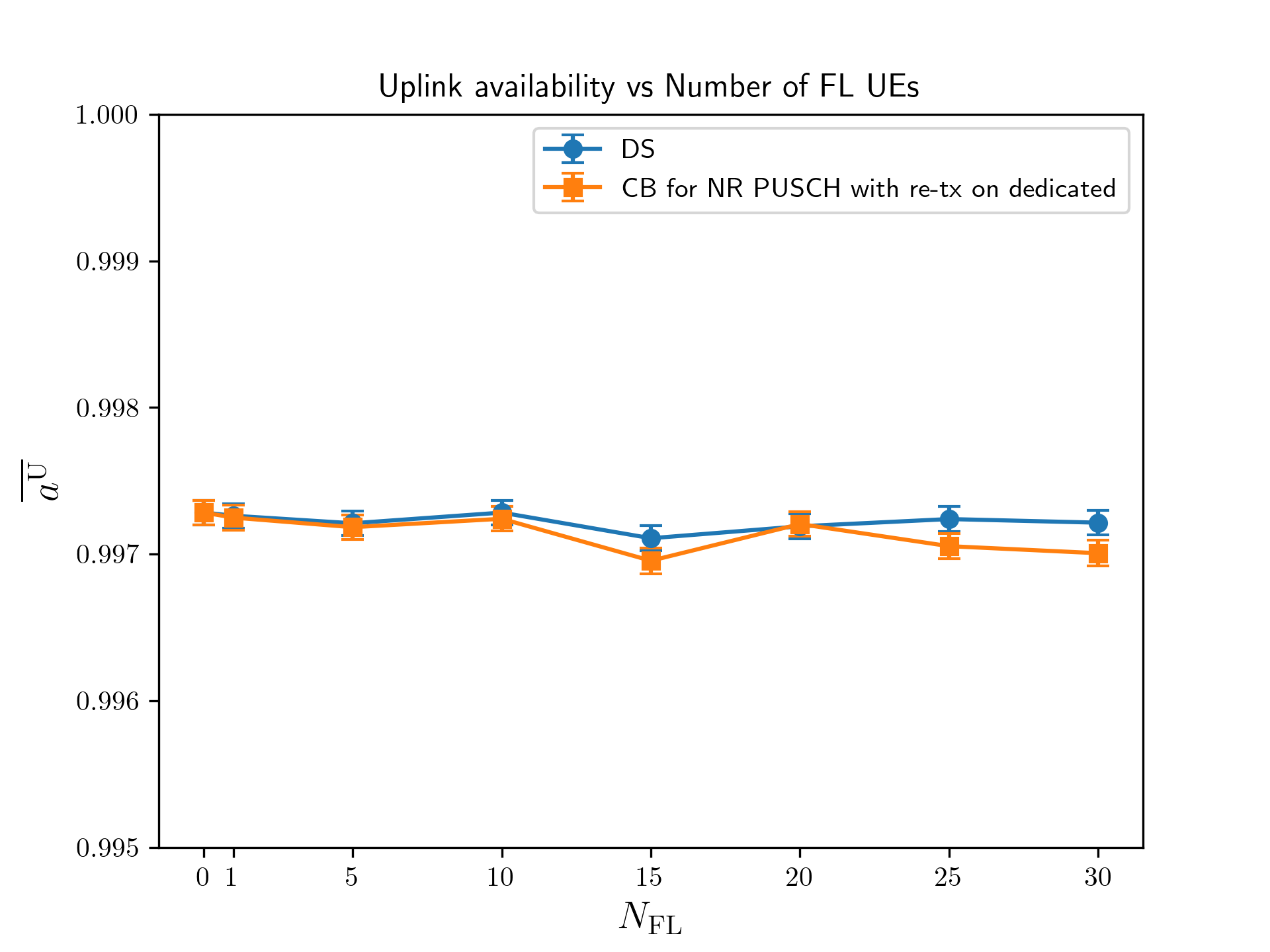}
	\label{fig:uplink_availability_12kB}}
    \subfloat[Average downlink availability of 10 \gls{urllc} \glspl{ue}.]
    {\includegraphics[width=0.5\linewidth]{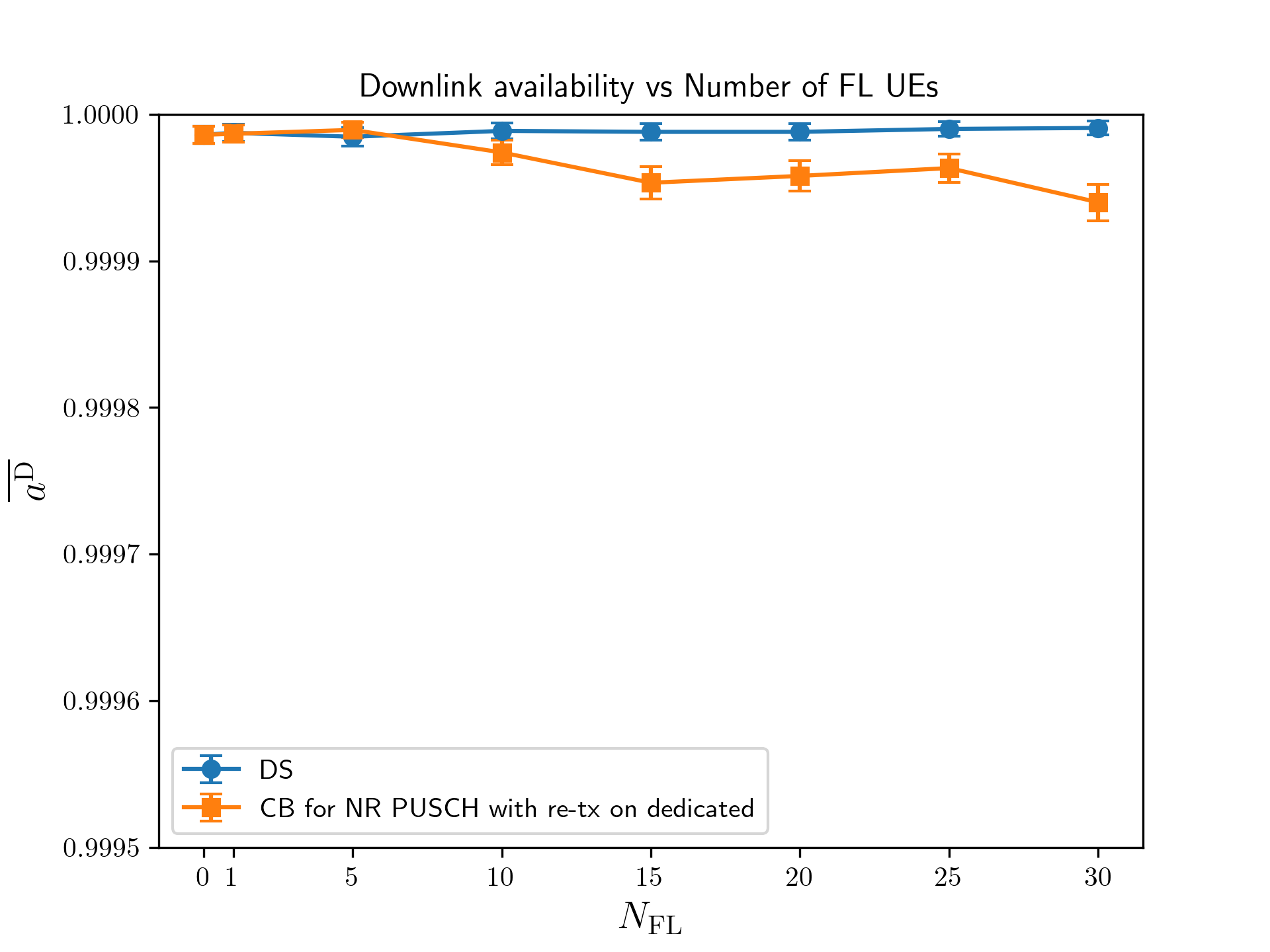}
	\label{fig:downlink_availability_12kB}}
\caption{Average uplink/downlink availability as a function of $N_{\rm FL}$, by comparing the two cases where \gls{fl} \glspl{ue} transmit/receive a model of 2 MB (i.e., the worst case) and are scheduled (i) via \gls{ds} or (ii) by means of the considered \gls{cb} for \gls{nr} \gls{pusch} with retransmissions on dedicated resources.}
\label{fig:availability}
\end{figure*}

Finally, Fig.~\ref{fig:availability} shows the \gls{urllc} uplink/downlink availability as a function of $N_{\rm FL}$, by comparing the two cases where \gls{fl} \glspl{ue} transmit/receive a model of 2 MB (i.e., the worst case) and are scheduled (i) via \gls{ds} or (ii) by means of the considered \gls{cb} design with retransmissions on dedicated resources. As expected, the uplink/downlink availability remain quite stable when increasing the number of \gls{fl} \glspl{ue} due to the higher priority of the \gls{urllc} traffic. However, the considered \gls{cb} design slightly (0.05\%) decreases the uplink/downlink availability because \gls{fl} retransmissions have higher priority w.r.t first transmissions of \gls{urllc} \glspl{ue} (thus the uplink/downlink delay bound is less easily met). %It is then worth highlighting that the downlink availability respects most of the \gls{iiot} requirements.

%%%%%%%%%%%%%%%%%%%%%%%%%%%%%%%%%%%%%%%%%%%%%%
%%%%%%%%%%%%%%%%%%%%%%%%%%%%%%%%%%%%%%%%%%%%%%
\section{Conclusions and Future Work}
\label{sec:conclusions}
In this paper, we presented a \gls{cb} \gls{5g} \gls{nr} framework for the \gls{pusch} transmissions of \gls{fl} \glspl{ue} in an \gls{iiot} scenario which also includes \gls{urllc} \glspl{ue}. By means of near-product \gls{3gpp}-compliant network simulations, we showed that the considered \gls{cb} for \gls{nr} \gls{pusch} design provides benefits over \gls{ds} for \gls{fl} upload/download times in case of sufficiently small model sizes (up to 12/16 kB).
%, which are not compatible with real dimensions \cite{howard2017mobilenets}, 
%This is an interesting result because, for \gls{fl} applications requiring larger model sizes, compression schemes must be employed at the application layer to reduce the actual model sizes and unleash the gain provided by the considered \gls{cb} for \gls{nr} \gls{pusch}. Additionally, such a contention design scales well with an increasing number of \glspl{ue} %(that is one of the trends of the Industry 4.0 paradigm) 
Additionally, such a \gls{cb} design scales well with an increasing number of \glspl{ue} and does not meaningfully degrade the performance of the \gls{urllc} traffic (uplink/downlink availability impacted at most by 0.05\%). However, for larger model sizes, \gls{ds} shows much better robustness of performance and scalability, and gains with \gls{cb} for \gls{nr} \gls{pusch} are not present because the size of the \gls{cb} allocations shrinks very quickly even for relatively low collision probabilities (close to 12\% at maximum), thereby leading to longer transfer times. 

The study also opens other interesting research trends. %aspects to be further evaluated when considering the possible gains of using \gls{cb} transmissions for mobile networks over the legacy \gls{ds} approach. 
For example, additional analyses could add the comparison with other \gls{sps} mechanisms (e.g., configured grant), identify the proper metrics that the network could monitor to optimize the \gls{cb} allocation (e.g., network load), consider more complex \gls{cb} design (e.g., reserving multiple \gls{cb} allocations per different sets of \glspl{ue}), or offload \gls{cb} by only mapping specific traffic types (e.g., information with low-reliability requirements).

\bibliographystyle{unsrt}
\bibliography{main.bib}

\begin{IEEEbiography}[{\includegraphics[width=1in,height=1.25in,clip,keepaspectratio]{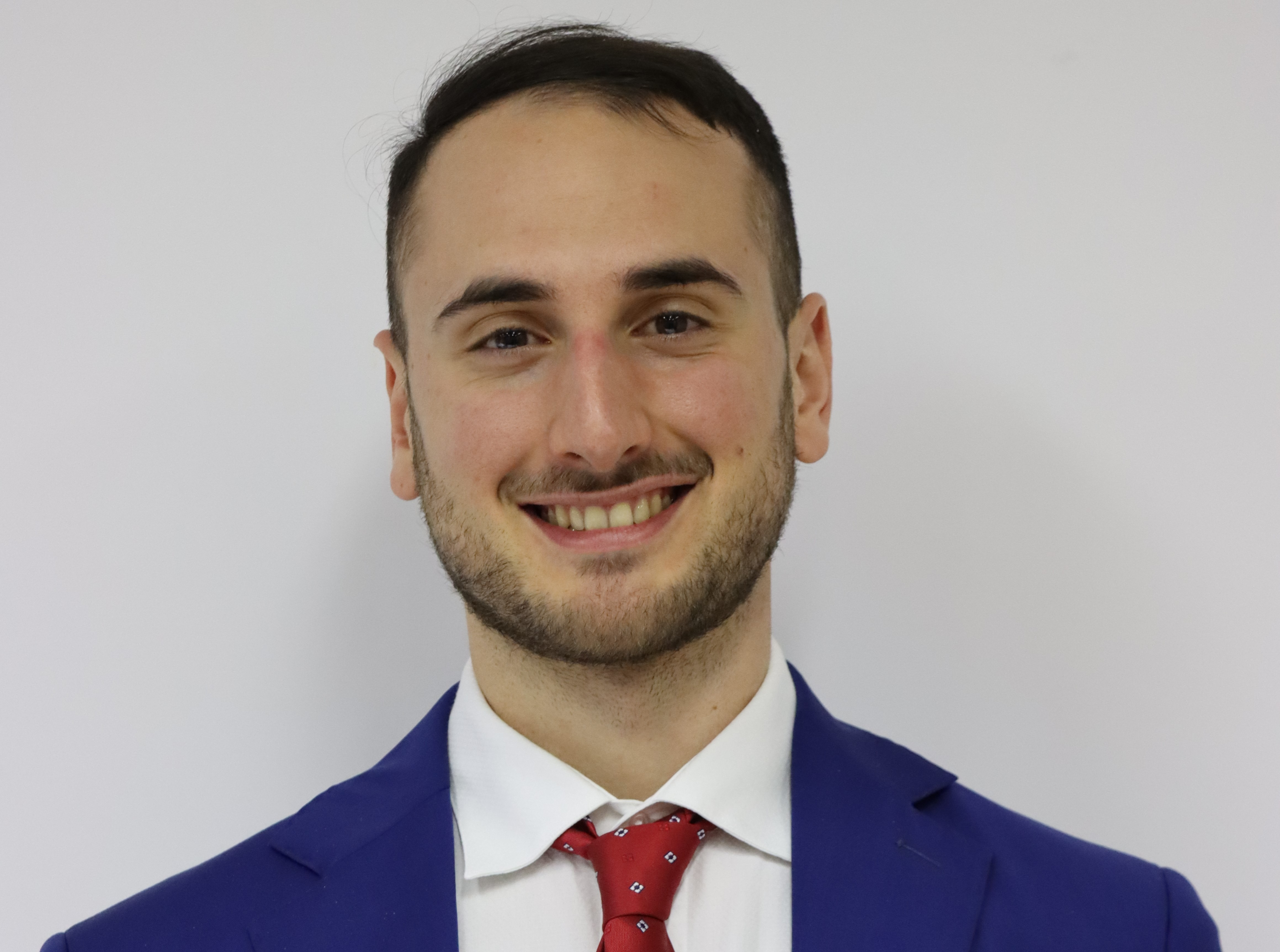}}]{Giampaolo Cuozzo} (M'2023) received a B.Sc with honors in electronics and telecommunications engineering and an M.Sc with honors in telecommunications engineering from the University of Bologna, in 2017 and 2019, respectively. He is currently pursuing a Ph.D. in electronics, telecommunications, and information technologies engineering (ETIT) at the University of Bologna. His research activity is focused on the study, development and validation of wireless networks for the Industrial Internet of Things, with a particular focus on MAC protocols for THz-based systems and scheduling optimization algorithms for 5G NR networks. He is also involved in experimental activities that exploit current wireless technologies, like 5G, LoRa, Zigbee and NB-IoT. 
\end{IEEEbiography}

\begin{IEEEbiography}[{\includegraphics[width=1in,height=1.25in,clip,keepaspectratio]{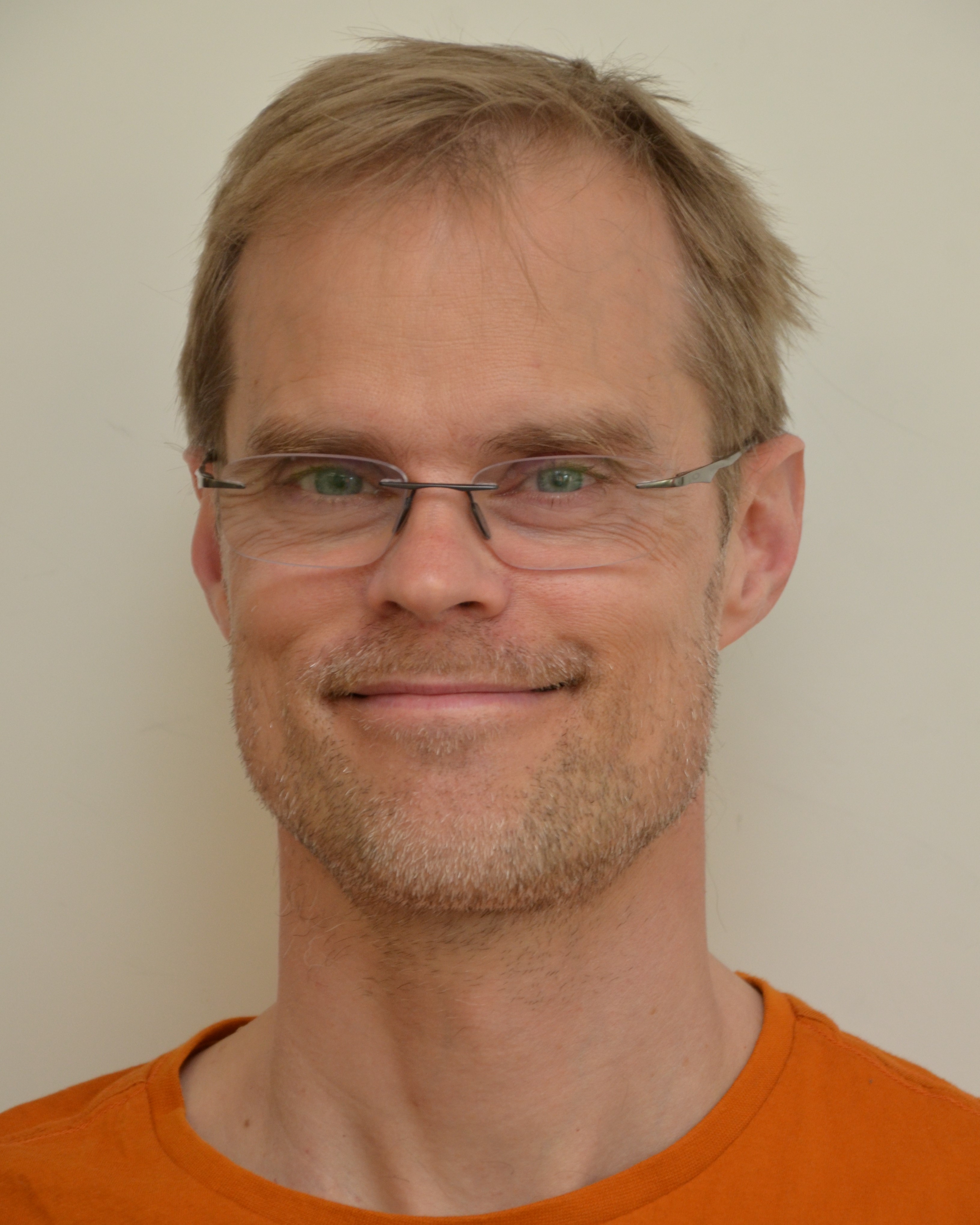}}]{Jonas Pettersson} received his M.Sc. in engineering physics from Umeå University in 1997 and joined Ericsson Research in Luleå the same year. He is currently working as Master Researcher in the protocol and end-to-end performance group. He has mainly worked in the areas of quality of experience, radio resource management, and scheduling for 3G, 4G, 5G, and now 6G. His present interests are time critical communication in 5G and medium access control for 6G. 
\end{IEEEbiography}

% was born in Greenwich Village, New York, NY, USA in 
% 1977. He received the B.S. and M.S. degrees in aerospace engineering from 
% the University of Virginia, Charlottesville, in 2001 and the Ph.D. degree in 
% mechanical engineering from Drexel University, Philadelphia, PA, in 2008.

% From 2001 to 2004, he was a Research Assistant with the Princeton Plasma 
% Physics Laboratory. Since 2009, he has been an Assistant Professor with the 
% Mechanical Engineering Department, Texas A{\&}M University, College Station. 
% He is the author of three books, more than 150 articles, and more than 70 
% inventions. His research interests include high-pressure and high-density 
% nonthermal plasma discharge processes and applications, microscale plasma 
% discharges, discharges in liquids, spectroscopic diagnostics, plasma 
% propulsion, and innovation plasma applications. He is an Associate Editor of 
% the journal \emph{Earth, Moon, Planets}, and holds two patents. 

% Dr. Author was a recipient of the International Association of Geomagnetism 
% and Aeronomy Young Scientist Award for Excellence in 2008, and the IEEE 
% Electromagnetic Compatibility Society Best Symposium Paper Award in 2011. 

\begin{IEEEbiography}[{\includegraphics[width=1in,height=1.25in,clip,keepaspectratio]{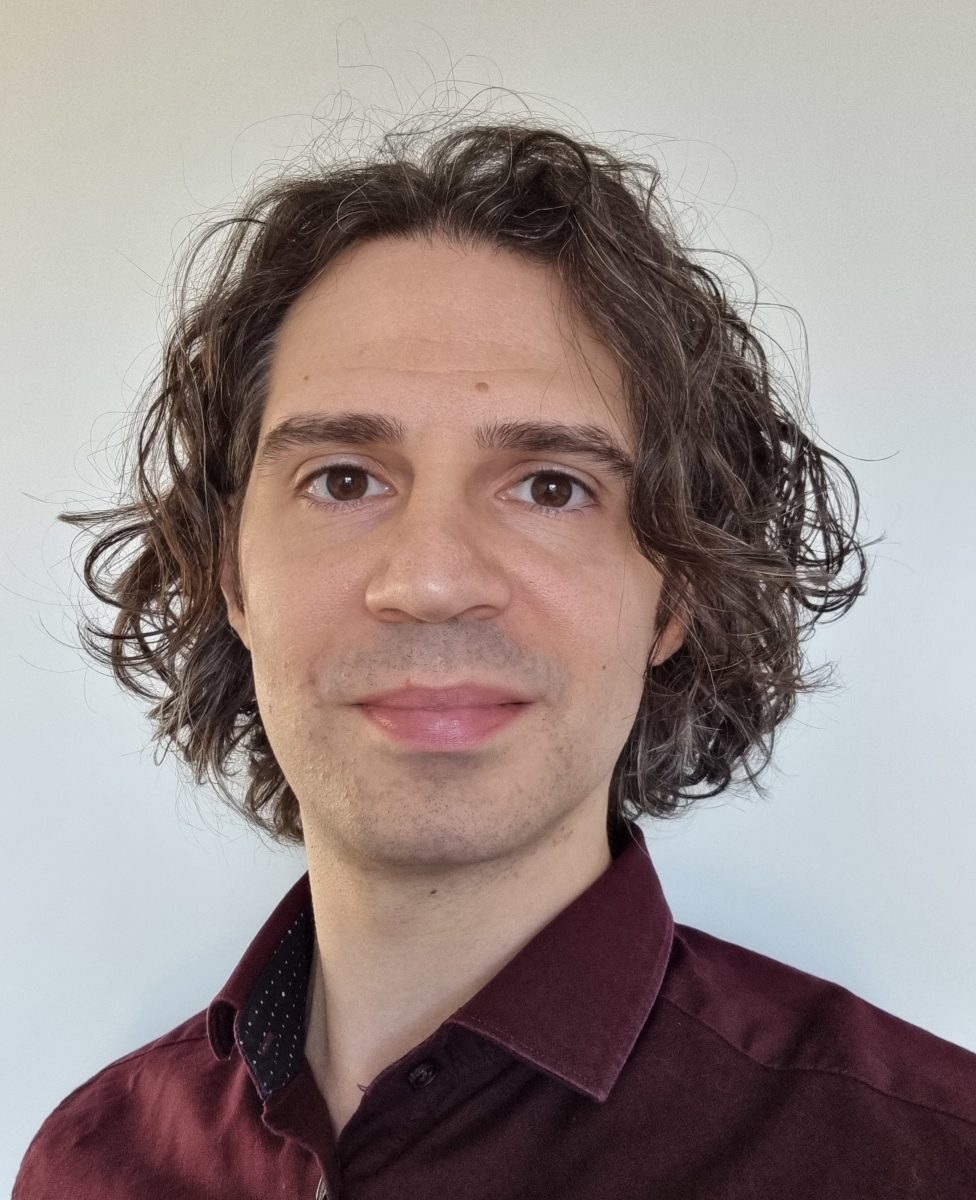}}]{Massimo Condoluci} is a Senior Researcher at Ericsson Research. He joined Ericsson in 2018 and he has worked on network exposure in 5G systems for automotive use cases, contributing to 3GPP standardization and to 5G Automotive Association (5GAA). He now focuses on beyond 5G mobile network architecture and protocols. From 2015 to 2017, he was a Research Associate at the Centre for Telecommunications Research (CTR), King's College London, UK, working on fixed-mobile convergence and radio-access optimization for haptic communications. He received the Ph.D. degree in information technology in 2016 from the University Mediterranea of Reggio Calabria, Italy, focusing on access optimization for IoT traffic and resource allocation for multicasting. From the same university, he received the M.Sc. and B.Sc. degrees in telecommunications engineering in 2011 and 2008, respectively. 
\end{IEEEbiography}

% \begin{IEEEbiography}[{\includegraphics[width=1in,height=1.25in,clip,keepaspectratio]{a3.png}}]{Third C. Author, Jr.} (M'87) received the B.S. degree in mechanical 
% engineering from National Chung Cheng University, Chiayi, Taiwan, in 2004 
% and the M.S. degree in mechanical engineering from National Tsing Hua 
% University, Hsinchu, Taiwan, in 2006. He is currently pursuing the Ph.D. 
% degree in mechanical engineering at Texas A{\&}M University, College 
% Station, TX, USA.

% From 2008 to 2009, he was a Research Assistant with the Institute of 
% Physics, Academia Sinica, Tapei, Taiwan. His research interest includes the 
% development of surface processing and biological/medical treatment 
% techniques using nonthermal atmospheric pressure plasmas, fundamental study 
% of plasma sources, and fabrication of micro- or nanostructured surfaces. 

% Mr. Author's awards and honors include the Frew Fellowship (Australian 
% Academy of Science), the I. I. Rabi Prize (APS), the European Frequency and 
% Time Forum Award, the Carl Zeiss Research Award, the William F. Meggers 
% Award and the Adolph Lomb Medal (OSA).
% \end{IEEEbiography}

\EOD

\end{document}